\documentclass[a4paper,cite,11pt,psnfss]{article}

\usepackage[T1]{fontenc}
\usepackage[latin1]{inputenc}
\usepackage{comment}
\usepackage{ulem}
\usepackage{adjustbox}

\usepackage{framed,epsfig,pgfplots,amsmath}
\usepackage{tikz}
\usepackage{tikz-feynhand}
\usetikzlibrary{snakes,arrows,shapes,positioning,automata,backgrounds,calc,er,patterns}
\renewcommand{\arraystretch}{1.5}

\usepackage[usenames,dvipsnames,tree]{pstricks}

\usepackage{graphicx,color,pst-plot}

\usepackage{latexsym,amsmath,amsfonts,amssymb,amstext,mathrsfs,upgreek}

\allowdisplaybreaks[1]

\begin{document}

\begin{titlepage}
\begin{flushright}
%
\end{flushright}
\begin{flushright}
\end{flushright}

\vfill

\begin{center}
{\Large\bf The three-loop QED contributions to the $g-2$ of charged leptons with two internal fermion loops and a class of Kamp\'e de F\'eriet series}

\bigskip
\bigskip

\vfill

{\bf B. Ananthanarayan$^a$, Samuel Friot~$^{b,c}$ and Shayan Ghosh~$^d$}\\[1cm]
{$^a$ Centre for High Energy Physics, Indian Institute of Science, \\
Bangalore-560012, Karnataka, India}\\[0.5cm]
{$^b$ Universit\'e Paris-Saclay, CNRS/IN2P3, IJCLab, 91405 Orsay, France } \\[0.5cm]
{$^c$ Univ Lyon, Univ Claude Bernard Lyon 1, CNRS/IN2P3, IP2I Lyon,\\
 UMR 5822, F-69622, Villeurbanne, France}\\[0.5cm]
{$^d$ Helmholtz-Institut f\"ur Strahlen- und Kernphysik \& Bethe Center for Theoretical Physics, Universit\"at Bonn, D-53115 Bonn, Germany} \\
\end{center}
\vfill

\begin{abstract}
The three-loop QED mass-dependent contributions to the $g-2$ of each of the charged leptons with two internal closed fermion loops, sometimes called $A^{(6)}_3\left(\frac{m_1}{m_2}, \frac{m_1}{m_3}\right)$ in the $g-2$ literature, is revisited using the Mellin-Barnes (MB) representation technique. Results for the muon and $\tau$ lepton anomalous magnetic moments $A^{(6)}_{3,\mu}$ and $A^{(6)}_{3,\tau}$, which were known as series expansions in the lepton mass ratios up to the first few terms only, are extended to their exact expressions. The contribution to the anomalous magnetic moment of the electron $A^{(6)}_{3,e}$ is also explicitly given in closed form. In addition to this, we show that the different series representations derived from the MB representation collectively converge for all possible values of the masses. Such unexpected behavior is related to the fact that these series bring into play double hypergeometric series that belong to a class of Kamp\'e de F\'eriet series which we prove to have the same simple convergence and analytic continuation properties as the Appell $F_1$ double hypergeometric series.

\end{abstract}

\end{titlepage}

\section{Introduction}

The Mellin-Barnes (MB) representation method, a well-known computational tool of perturbative quantum field theory, can be used to derive series representations of Feynman diagrams and related quantities in terms of multiple hypergeometric series. In general, once the Feynman diagram or quantity of interest has been represented by a multi-fold MB integral, a standard residue calculation shows that several of such series representations, converging in different parts of the parameter space, can be derived and these series are, as a rule, analytic continuations of one another (see \cite{Friot:2011ic} for a systematic exposition of the 2-fold case). However, even at the level of 2-fold MB representations the convergence regions of these analytic continuations do not collectively cover, in general, the whole parameter space of the computed quantity. This implies that one has to find alternative (and sometimes non trivial) analytic continuation methods in order to obtain analytic expressions valid in the particular regions of the parameter space where none of the series derived from the standard residue computation of the MB representation can be used. We call these inaccessible regions the "white regions" in what follows.

A well-known example involving triple series is the two-loop massive sunset Feynman diagram. In \cite{Berends:1993ee}, two different triple series representations of the latter, derived from its 3-fold MB representation have been given in closed form as combinations of Lauricella $F_C^{(3)}$ triple series, and two others can also be obtained, either from the MB representation or by using the invariance of the $F_C^{(3)}$ series under any permutation of its variables. These four series representations, analytic continuations of one another, converge in different regions of the 3-dimensional $\left(\frac{p^2}{m_3^2},\frac{m_1^2}{m_3^2},\frac{m_2^2}{m_3^2}\right)$ parameter space of the sunset diagram (where $p$ is the external momentum and the $m_i$ are the masses of the involved particles), but there remains a white region, which includes regions of phenomenological interest, that cannot be reached by any of them. We have shown in \cite{Ananthanarayan:2019icl} how one can analytically continue some of these series to get new series representations of the sunset diagram that can be used to analytically evaluate the latter in several important parts of its white region. 

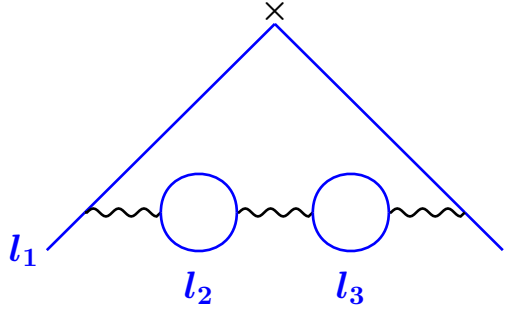
\begin{figure}
\begin{center}
\begin{tikzpicture}[scale=1]
\begin{feynhand}
\vertex (a) at (1,1);
\vertex (b) at (1.5,1.5);
\vertex (c) at (4,4);
\vertex (d) at (6.5,1.5);
\vertex (e) at (7,1);
\vertex (f) at (2.5,1.5);
\vertex (g) at (3.5,1.5);
\vertex (h) at (4.5,1.5);
\vertex (i) at (5.5,1.5);
\propag [plain,blue] (a) to (b);
\propag [plain,blue] (b) to (c) node[above=-4pt] {\black{\large{$\boldsymbol{\times}$}}};
\propag [plain,blue] (c) to (d);
\propag [plain,blue] (d) to (e);
\propag [boson] (b) to (f);
\propag [boson] (g) to (h);
\propag [boson] (i) to (d);
\propag[plain,blue] (f) to [half left,looseness=1.75] (g);
\propag[plain,blue] (f) to [half right,looseness=1.75] (g);
\propag[plain,blue!100!black] (h) to [half left,looseness=1.75] (i);
\propag[plain,blue!100!black] (h) to [half right,looseness=1.75] (i);
\node at (0.7,1) {\textcolor{blue}{\Large $\boldsymbol{l_1}$}};
\node at (3,0.5) {\textcolor{blue}{\Large $\boldsymbol{l_2}$}};
\node at (5,0.5) {\textcolor{blue!100!black}{\Large $\boldsymbol{l_3}$}};
\end{feynhand}
\end{tikzpicture}
\end{center}
\caption{The 3-loop QED Feynman diagram corresponding to $A^{(6)}_3(m_1/m_2,m_1/m_3)$.\label{g-2_3loop}} 
\end{figure}

In this paper, we go further on our exploration of the analytic continuation properties of Feynman diagrams and related quantities by revisiting what is possibly the simplest class of QED contributions to the anomalous magnetic moment of each of the charged leptons that can be represented by a 2-fold MB integral. These three-loop QED mass-dependent contributions with two internal closed fermion loops (see Figure \ref{g-2_3loop} for the corresponding Feynman diagram), often denoted $A^{(6)}_3(m_1/m_2,m_1/m_3)$ in the $g-2$ literature, can then involve at most double hypergeometric series, and we show that they have an unexpected behavior. Indeed, the analytic continuation properties that these $g-2$ contributions satisfy are, surprisingly, the converse of what one faces when one deals with, for instance, the sunset diagram case because the MB representation of $A^{(6)}_3(m_1/m_2,m_1/m_3)$ does not give rise to any white region. This interesting result has encouraged us to probe what is special about the double hypergeometric series involved in our final expressions. Studying the specific form of these series, we observe that they have the same simple convergence and analytic continuation properties as the Appell $F_1$ double hypergeometric series. Furthermore, we show that the latter and the former both belong to a class of Kamp\'e de F\'eriet series for which we prove, from their MB representation, the absence of white regions.

As another motivation for studying these particular $g-2$ contributions it should be noted, and as emphasized in \cite{Jegerlehner:2017gek}, that in contrast to all other three loop QED contributions to the muon anomalous magnetic moment, $A_{3,\mu}^{(6)}(m_\mu/m_\tau,m_\mu/m_e)$ is the only one whose exact analytic form has not been derived so far. Results were first presented in \cite{Czarnecki:1998rc} in terms of the first few terms of a series expansion in powers and logarithms of the mass ratios, using large-momentum, heavy mass and eikonal expansions techniques. These results have then been checked and extended in \cite{Friot:2005cu} using the MB representation method. In the present paper, we have derived them in their entirety and present their exact expressions, in terms of generalised hypergeometric  and Kamp\'e de F\'eriet double hypergeometric series.
In the case of the electron, we have not been able to find any analytic result for these contributions in the $g-2$ literature, although some numerical evaluations of these have been given, for instance in \cite{Passera:2007fk}. We will show in the following that the exact expression of $A_{3,e}^{(6)}(m_e/m_\tau,m_e/m_\mu)$ has a simple and compact form.
The $\tau$ lepton case is more intricate and has been considered a long time ago in \cite{Narison:1977jc}. The latter reference gives, to our knowledge, the only available non-numerical result for $A_{3,\tau}^{(6)}(m_\tau/m_\mu,m_\tau/m_e)$. The result of \cite{Narison:1977jc} corresponds to the leading term in the double series expansion of the exact expression which we will present in the following. The numerical evaluation of this leading term, presented in \cite{Narison:2001jt}, does not agree with the numerical evaluation of $A_{3,\tau}^{(6)}(m_\tau/m_\mu,m_\tau/m_e)$ given in \cite{Passera:2007fk}. We show here that this mismatch can be solved once one adds some sub-leading terms to the expression of \cite{Narison:1977jc,Narison:2001jt}.

In view of all the considerations spelt out in the foregoing, we now give the outline of this paper. In Section~\ref{Sec:QED}, a short review of the QED contributions to the anomalous magnetic moment of charged leptons is given. In Section \ref{Sec:QED2} we present the MB representation for the  three-loop contribution to $g-2$ coming from the Feynman diagrams of Figure \ref{g-2_3loop}, and then calculate it for the cases with external electron, muon and $\tau$ lepton. Detailed expressions for each of these are listed in the Appendix. In Section \ref{Sec:checks}, we present the checks of our formulas and give a brief numerical analysis, and in Section \ref{MBAC} we discuss the analytic continuation properties of the class of Kamp\'e de F\'eriet series mentioned above. We conclude with Section \ref{concl}, where a short discussion of the results and future work are presented.

\section{Short QED literature review \label{Sec:QED}}

The anomalous magnetic moment of the charged leptons is defined as $a_l \equiv (g_l-2)/2$, where $g_l$ is the Land\'e factor and $l=e, \mu, \tau$. In the Standard Model, contributions to $a_l$ arise from electroweak and strong processes. 
The anomalous magnetic moment of charged leptons has a  distinguished place in elementary particle physics. Historically, the electron anomalous magnetic moment has been among the most important tests of quantum electrodynamics (QED). During the last decade, a persistent discrepancy between the Standard Model theoretical predictions and experimental results in the case of the muon, now reaching a $3.5 \ \sigma$ level, has spurred a new experiment at Fermilab, from which a $5\ \sigma$ deviation could well be obtained in the near future \cite{Knecht:2018nhq}. Due to these reasons, over the decades a huge amount of theoretical effort has been devoted to computing this quantity (see \cite{Jegerlehner:2017gek} for a recent and comprehensive review in the muon case), with a great deal of activity being directed at computing hadronic contributions to it. We recall that, due to the larger mass of the muon, it is usually accepted that the muon anomalous magnetic moment is more sensitive to new physics than the electron. The very short life time of the $\tau$ lepton prevents the experimental measurement of its anomalous magnetic moment, which explains why the theoretical study of the latter is less well developed (see \cite{Eidelman:2007sb} for a review of the $\tau$ lepton case).
The best experimental limits are $-0.052<a_\tau<0.013\  (95\% \text{CL})$ \cite{Abdallah:2003xd}.

The QED contributions to the anomalous magnetic moment of the charged leptons can be expressed perturbatively as

\begin{align}
    a_l^{\text{QED}} = A_{1,l} + A_{2,l} \left( \frac{m_l}{m_{l'}} \right) + A_{2,l} \left( \frac{m_l}{m_{l''}} \right) + A_{3,l} \left( \frac{m_l}{m_{l'}}, \frac{m_l}{m_{l''}} \right)
\end{align}
with
\begin{align}
    A_{i,l} = A_{i,l}^{(2)} \left( \frac{\alpha}{\pi} \right) + A_{i,l}^{(4)} \left( \frac{\alpha}{\pi} \right)^2 + A_{i,l}^{(6)} \left( \frac{\alpha}{\pi} \right)^3 + ... 
\end{align}
where $A_{i,l}^{(2n)}$ is the sum of the $n^{\text{th}}$ loop contributions, and  $A_{2,l}^{(2)} = A_{3,l}^{(2)} = A_{3,l}^{(4)} = 0$.

The $A_{1,l}$ are mass independent and thus equivalent for all three lepton flavours. Up to $\mathcal{O}(\alpha^3)$, these contributions are known in closed analytic form, while $A_1^{(8)}$ and higher-loop contributions are only known numerically. That $A_1^{(2)} = 1/2$ has long been known \cite{Schwinger:1948iu}, as has the value of $A_1^{(4)}$ \cite{Karplus:1950zzb, Petermann:1957hs, Sommerfield:1957zz}. A numerical value of $A_1^{(6)}$ is given in \cite{Kinoshita:1995ym}, and after work spanning several decades in calculating the various diagrams that constitute $A_1^{(6)}$ 
(e.g. \cite{Barbieri:1974dg, Barbieri:1975qf, Levine:1979uz, Laporta:1991zw, Laporta:1993qw, Laporta:1994yi, Laporta:1995vp}), the calculation was finalised and the exact close form analytic result was presented in \cite{Laporta:1996mq}. Purely numerical calculations for $A_1^{(8)}$ can be found in \cite{Kinoshita:1981wm, Hughes:1999fp, Kinoshita:2005zr, Aoyama:2007dv, Aoyama:2007mn, Aoyama:2012wj, Aoyama:2014sxa}, and a result to 1100 digit precision, accompanied by a semi-analytic fit to the result, is presented in \cite{Laporta:2017okg}. A numerical value for $A_1^{(10)}$ is given in \cite{Aoyama:2012wj, Aoyama:2014sxa}.

The mass dependent terms $A_{2,l}^{(4)}$ have been calculated as a series expansion in the small mass ratio \cite{Petermann:1957ir, Samuel:1990qf, Li:1992xf, Czarnecki:1998rc}, as an exact result for small mass ratios \cite{Elend:1966}, and finally as a closed form analytic result for all values of the mass ratio \cite{Passera:2004bj}. As the diagrams under consideration in this work can contribute at the three-loop level when two of the three involved leptons have the same flavor, we discuss $A_{2,l}^{(6)}$ in greater detail below. The fourth and higher loop results were known primarily numerically, e.g. \cite{Aoyama:2012wk,Aoyama:2012wj, Aoyama:2014sxa}, although some analytic results are also available, e.g. \cite{Kurz:2016bau, Kurz:2013exa, Kataev:2012kn, Aguilar:2008qj, Solovtsova:2019act}.

One of the earliest calculations of $A_{2,\mu}^{(6)}$ was done in  \cite{Kinoshita:1967}, where an expansion including some of the leading log and analytic terms was presented. This expansion was extended in \cite{Lautrup:1969fr}, and a result for Figure~\ref{g-2_3loop}, with $l_1=\mu$ and $l_2=l_3=e$ was given with log contributions up to order $(m_\mu/m_e)^0$ included. In \cite{Billi:1972bn, Barbieri:1972ek, Barbieri:1974pk, Barbieri:1974dg} calculations and results for some diagrams making up $A_{2,\mu}^{(6)}$ and $A_{2,e}^{(6)}$ are presented. In \cite{Barbieri:1974nc} analytic results for vacuum polarization contributions to $A_{2,\mu}^{(6)}$ and $A_{2,e}^{(6)}$, up to order $(m_\mu/m_e)^0$, are given. Expressions for all the individual graphs (such as for Figure~\ref{g-2_3loop}) are however not presented. \cite{Lautrup:1977tc} completes the calculation of all the $\log(m_\mu/m_e)$ terms by computing light-by-light scattering diagrams, numerical estimates for which are to be found in \cite{Kinoshita:1988yp, Samuel:1992ub}. The electron light-by-light scattering graph contributions to $A_{2,\mu}^{(6)}$ are calculated in \cite{Laporta:1992pa}, and the expansion presented there is extended to higher order in \cite{Kuhn:2003pu}. 
\cite{Samuel:1990qf} continues the expansion given in \cite{Barbieri:1974nc} to order $(m_\mu/m_e)^0$. In \cite{Laporta:1993ju} close form analytic results for the vacuum polarization diagrams contributing to $A_{2,\mu}^{(6)}$ are given, as are expansions in mass ratios (up to certain order) that can be used to calculate the equivalent contributions to $A_{2,e}^{(6)}$ and $A_{2,\tau}^{(6)}$. The exact expression given in \cite{Laporta:1993ju} is also expanded in the small mass ratio in \cite{Passera:2006gc}, but to a higher order than in the former, and which is then used to calculate and compare with numerically evaluated values of $A_{2,e}^{(6)}$.

The three loop contribution that we are primarily concerned with in this work is the lowest order non-zero term consisting of three masses (or two mass ratios), $A_{3,l}^{(6)}$. In \cite{Samuel:1990qf}, an integral representation is given for $A_{3,l}^{(6)}$ and numerically evaluated for $l=\mu$ (we have noted that Eqs.(32) and (33) of \cite{Samuel:1990qf} do not match numerically because of a 2 overall factor that is missing in the r.h.s. of Eq.(32)). Another numerical evaluation of $A_{3,\mu}^{(6)}$ is given in \cite{Laporta:1993ju}. For $A_{3,\mu}^{(6)}$, \cite{Czarnecki:1998rc} gives an expansion up to the first few terms, based on asymptotic and eikonal methods, while in \cite{Friot:2005cu} these results are slightly extended, using the  Mellin-Barnes technique. In \cite{Aoyama:2012wk}, results for QED contributions to muon $g-2$ up to the tenth order are given, which includes numerical values for $A_{3,\mu}^{(6)}$ and $A_{2,\mu}^{(6)}$. The integral of \cite{Samuel:1990qf} is evaluated for the electron case, and a numerical value for $A_{3,e}^{(6)}$ is given in \cite{Passera:2006gc}. Numerical results for the same, as well as for $A_{2,e}^{(6)}$, but based on an older set of mass value inputs, is given in \cite{Kurz:2013exa}. And \cite{Aoyama:2012wj} gives numerical results for QED contributions to electron $g-2$ up to the tenth order, including values for $A_{3,e}^{(6)}$ and $A_{2,e}^{(6)}$. \cite{Narison:2001jt} reviews and updates all contributions to the muon and $\tau$ lepton $g-2$ up to its date of publication, and based on the results of \cite{Narison:1977jc}, gives an expression for $A_{3,\tau}^{(6)}$ as an expansion up to its leading term, which corresponds to our $R_{\{3,1\}}$ result, see Section \ref{QEDtau}, Table \ref{Table:TauResults}. Its numerical value is $2.75316$ \cite{Narison:2001jt} and can be compared to the numerical value for $A_{3,\tau}^{(6)}$ 
given in \cite{Samuel:1990su}: $1.679$. The numerical result for $A_{3,\tau}^{(6)}$ calculated in \cite{Passera:2006gc} (see also \cite{Eidelman:2007sb}) is 3.34797, which therefore disagrees with those of \cite{Narison:2001jt, Samuel:1990su} but agrees with our own calculated value as shown later on in this work. Since the result of \cite{Samuel:1990su} is precisely half of the one of \cite{Passera:2006gc}, we suppose that, as mentioned above, the factor of 2 missing in the r.h.s. of Eq.(32) of \cite{Samuel:1990qf} is at the origin of this discrepancy, since the authors of \cite{Samuel:1990qf} are also those of \cite{Samuel:1990su}. Concerning the discrepancy with \cite{Narison:1977jc, Narison:2001jt}, as already said in the introduction, adding a few sub-leading terms allows to obtain agreement with \cite{Passera:2006gc}.

For a recent review of the theory and experimental status of the $g-2$, see \cite{Knecht:2018nhq} (\cite{KNECHT:2014bsa} is also useful). For a comprehensive review of QED contributions to all the leptons' $g-2$, see \cite{Laporta:2018dyv, Passera:2006gc}. A review of the muon $g-2$ is given in \cite{Logashenko:2018pcl, Jegerlehner:2017gek, Miller:2012opa, Miller:2007kk, Passera:2004bj}, and a review of contributions to $\tau$ lepton $g-2$ is presented in \cite{Eidelman:2007sb}.

\section{Three-loop QED contributions with two internal loops \label{Sec:QED2}}

We will now give the exact expressions of the contributions $A_{3,l}^{(6)}$ to the anomalous magnetic moments of the electron, muon and $\tau$ lepton coming from the Feynman diagram of Figure \ref{g-2_3loop} (and the symmetric diagram obtained by an exchange of the internal loops).

\begin{figure}[h]
\begin{center}
\includegraphics[height=5cm]{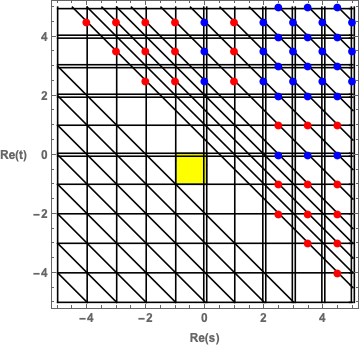}
\caption{Singular structure of the integrand of Eq.(\ref{MBQED})\label{singQED}. The red dots indicate points where the singularity has been cancelled, and the blue dots indicate points where the order of the singularity has been reduced, due to factors in the denominator of Eq.(\ref{MBQED}).}
\end{center}
\end{figure}

The MB representation of these contributions may be found in \cite{Friot:2005cu}. Defining $r_1\doteq m_{l_1}^2/m_{l_2}^2$ and $r_2\doteq m_{l_1}^2/m_{l_3}^2$, it reads:
\begin{multline}
\label{MBQED}
A_{3,l}^{(6)}(\sqrt{r_1},\sqrt{r_2})=\frac{\sqrt{\pi}}{8} \int \limits_{\boldsymbol{\gamma}+i\mathbf{R}^2}\!\!\frac{d s}{2i\pi}\wedge\frac{d t}{2i\pi}\; r_1^{-s}\; r_2^{-t} \;\; \Gamma(s)\Gamma(-s)\Gamma(t)\Gamma(-t)\Gamma(2-s)\Gamma(2-t) \\
\times\frac{\Gamma\left(\frac{1}{2}-s-t\right)\Gamma(1-s-t)\Gamma(2+s+t)}{\Gamma\left(\frac{5}{2}-s\right)\Gamma\left(\frac{5}{2}-t\right)\Gamma(3-s-t)}
\end{multline}
where $\gamma\doteq(\textrm{Re}(s),\textrm{Re}(t))\in]-1,0[\times]-1,0[$ (see the yellow region in Figure \ref{singQED}).

From the rules described in \cite{Friot:2011ic, Tsikh:1998} it is clear that we are in a so-called degenerate case $({\bf\Delta}=0)$, where several convergent series representations of the integral coexist, being, as a rule, analytic continuations of one another.

\begin{figure}[h]
\begin{center}
\includegraphics[height=7cm]{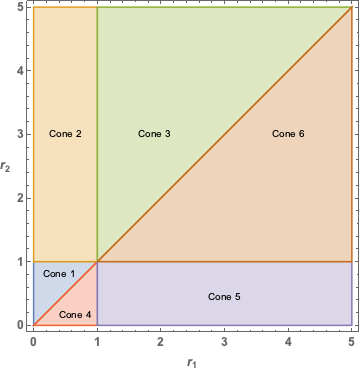} 
\caption{Convergence regions of the series representations of Eq.(\ref{MBQED}) labelled by their associated cone\label{region_conv}.}
\end{center}
\end{figure}

Since the MB integral is fully symmetric under the exchange of $r_1$ and $r_2$ (or $s$ and $t$) one can avoid about half of the calculations that would be necessary to perform in order to derive all the possible convergent series representations in the case of a non symmetrical integral. This symmetry, which comes from the symmetry of the Feynman diagram under the exchange of $l_2$ and $l_3$, is also reflected in the singular structure of the integrand (see Figure \ref{singQED}) and in the picture showing the convergence regions of the series representations in the first quadrant of the $(r_1,r_2)$-plane (see Figure \ref{region_conv}).

\begin{figure}[h]
\begin{center}
\includegraphics[width=4.5cm]{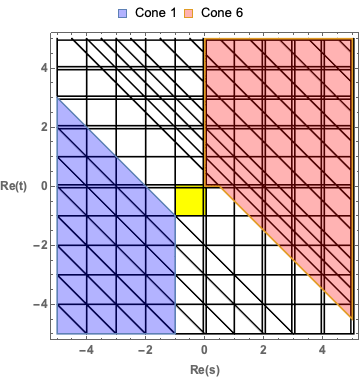} \hspace{0.2cm} \includegraphics[width=4.5cm]{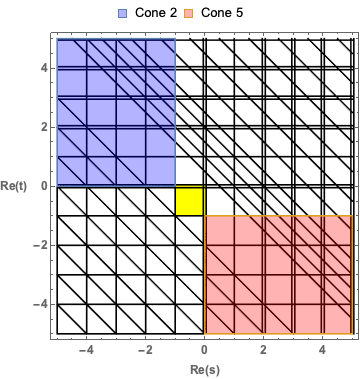} \hspace{0.2cm} \includegraphics[width=4.5cm]{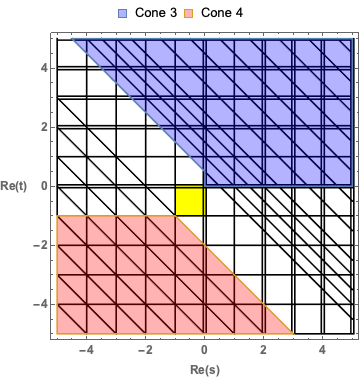}
\end{center}
\caption{The cones associated to the MB integral of Eq.(\ref{MBQED})\label{conesQED}}
\end{figure}

It is easy to find the different sets of residues (cones) associated to each convergent series (see \cite{Friot:2011ic} for details on the general procedure). There are six such cones, plotted in Figure \ref{conesQED} and, as just explained, the series representations associated to only three of them have to be computed (the blue cones), the others (coming from the red cones) being obviously derived from the latter by exchanging $r_1$ and $r_2$ in the final results.

We show in Figure \ref{region_conv} the convergence regions of the series representations deduced from each of the cones and, in Table \ref{Table:TwoMassCones}, we show to which of the cones the possible physical situations are associated. The mass independent case can be computed from any of the cones, as can be seen in Figure \ref{region_conv} .

\begin{table}
\renewcommand{\arraystretch}{2.0}
\begin{tabular}{ |c|c|c|c| } 
 \hline
 $l_1$ & $l_2$ & $l_3$ & Cones  \\ 
 \hline
 \hline 
    $m_e$ & $m_\tau$ & $m_\mu$ & 1 \\
    $m_e$ & $m_\mu$ & $m_\tau$ & 4 \\
    $m_e$ & $m_e$ & $m_\mu$ & 4,5 \\
    $m_e$ & $m_\mu$ & $m_e$ & 1,2 \\
    $m_e$ & $m_e$ & $m_\tau$ & 4,5 \\ 
    $m_e$ & $m_\tau$ & $m_e$ & 1,2 \\    
    $m_e$ & $m_\mu$ & $m_\mu$ & 1,4 \\
    $m_e$ & $m_\tau$ & $m_\tau$ & 1,4 \\
 \hline
\end{tabular} \hspace{7mm}
\begin{tabular}{ |c|c|c|c| } 
 \hline
 $l_1$ & $l_2$ & $l_3$ & Cones  \\ 
 \hline
 \hline
    $m_\mu$ & $m_\tau$ & $m_e$ & 2 \\
    $m_\mu$ & $m_e$ & $m_\tau$ & 5 \\
    $m_\mu$ & $m_\mu$ & $m_e$ & 2,3 \\
    $m_\mu$ & $m_e$ & $m_\mu$ & 5,6 \\
    $m_\mu$ & $m_\mu$ & $m_\tau$ & 4,5 \\
    $m_\mu$ & $m_\tau$ & $m_\mu$ & 1,2 \\
    $m_\mu$ & $m_e$ & $m_e$ & 3,6 \\
    $m_\mu$ & $m_\tau$ & $m_\tau$ & 1,4 \\
\hline
\end{tabular} \hspace{7mm}
\begin{tabular}{ |c|c|c|c| } 
 \hline
 $l_1$ & $l_2$ & $l_3$ & Cones  \\ 
 \hline
 \hline
    $m_\tau$ & $m_\mu$ & $m_e$ & 3 \\
    $m_\tau$ & $m_e$ & $m_\mu$ & 6 \\
    $m_\tau$ & $m_\tau$ & $m_\mu$ & 2,3 \\
    $m_\tau$ & $m_\mu$ & $m_\tau$ & 5,6 \\
    $m_\tau$ & $m_\tau$ & $m_e$ & 2,3 \\
    $m_\tau$ & $m_e$ & $m_\tau$ & 5,6 \\
    $m_\tau$ & $m_e$ & $m_e$ & 3,6 \\
    $m_\tau$ & $m_\mu$ & $m_\mu$ & 3,6 \\
 \hline
\end{tabular}
\caption{The different combinations of charged leptons and corresponding cones. $l_1$ is the external lepton (see Figure \ref{g-2_3loop}).}
\label{Table:TwoMassCones}
\end{table}

\subsection{The electron case (Cone 1)\label{QEDelectron} }

Let us begin with the simplest case, namely $A_{3,e}^{(6)}$: a muon and a tau in the internal loops and an electron on the external legs of the Feynman diagram shown in Figure \ref{g-2_3loop}, and its symmetric counterpart. We will see that the exact analytic expression of these contributions to the anomalous magnetic moment of the electron is more compact than the corresponding expressions for the muon and tau cases. As mentioned in the introduction, we have not been able to find any analytic result for this contribution in the $g-2$ literature.

A convergence analysis to be presented below shows that the series representation corresponding to the electron case comes from Cone 1 (or Cone 4 for the symmetric diagram) in Figure~\ref{conesQED}.
One can see on Figures~\ref{conesQED} and \ref{singQED} that this cone is simpler than Cone 2 and Cone 3, since there are only four different subsets of residues to compute and because there is no interference between the gamma functions of the numerator and those of the denominator in the integrand of Eq.(\ref{MBQED}).  With $m$ and $n$ any non-negative integer, the coordinates of the associated poles in Figure \ref{conesQED} are :
\begin{itemize}
\item Single series contributions: $(-2-m,0)$, $(-3-m,1)$.
\item Double series contributions: $(-4-m-n,2+m)$, $(-1-m,-1-n)$.
\end{itemize}

\subsubsection{Exact result}

The series representation of these contributions to the anomalous magnetic moment of the electron is expressed in closed form in terms of the generalized hypergeometric series ${}_4F_3$ and the Kamp\'e de F\'eriet double hypergeometric series $F^{2:3;3}_{2:2;2}$ and $F^{2:3;2}_{2:2;1}$, and the final expression reads 
\begin{equation}
A_{3,e}^{(6)}(\sqrt{r_1},\sqrt{r_2})=\frac{\sqrt{\pi}}{8}\sum_{i=1}^{4} R_{\{1,i\}}\ \hspace{5mm} \text{where} \hspace{5mm} r_1 = \frac{m_e^2}{m_\tau^2}, \quad r_2 = \frac{m_e^2}{m_\mu^2}.
\label{electrons_results}
\end{equation} 
The residues $R_{\{1,i\}}$ are given in Table~\ref{Table:ElectronResults} as well as their correspondence to the singular points listed above. As the expressions for some of the residues is lengthy, we have introduced some functions, $h_{\{1,j\}}$, in the table, whose explicit forms are given in the Appendix, see Eqs.(\ref{Ap_el1})-(\ref{Ap_el4}).

\begin{table}
\renewcommand{\arraystretch}{1.75}
\center
\begin{tabular}{ |c|c|c| } 
 \hline
 Singularity & Label & Residue \\ 
 \hline
 \hline
  $(-1-m,-1-n)$ & $R_{\{1,1\}}$ & $ \left[ \frac{\partial}{\partial \alpha} h_{\{1,1\}} \right]_{\alpha=0}$ \\ 
 $(-2-m,0)$ & $R_{\{1,2\}}$ & $ \left[ \left( \frac{2 \pi^2}{3} +  \frac{\partial^2}{\partial \alpha^2} \right) h_{\{1,2\}}  \right]_{\alpha=0}$ \\ 
 $(-3-m,1)$ & $R_{\{1,3\}}$ & $ \left[ \frac{\partial}{\partial \alpha} h_{\{1,3\}}  \right]_{\alpha=0}$ \\
 $(-4-m-n,2+m)$ & $R_{\{1,4\}}$ & $ \left[ \frac{\partial^2}{\partial \alpha^2}  h_{\{1,4\}}  \right]_{\alpha=0}$ \\ 
 
 \hline
\end{tabular}
\caption{Cone 1 results (simplified). Expressions for the $h_{\{1,i\}}$ can be found in Eqs.(\ref{Ap_el1})-(\ref{Ap_el4}) of the Appendix.}
\label{Table:ElectronResults}
\end{table}

Due to the presence of non-simple poles in the singularity structure of Eq.(\ref{MBQED}), many of its residues involve polygamma functions, which come from derivatives of the gamma function. As these polygamma terms arise from an application of the residue theorem, they are derivatives of gamma functions appearing in the $h_{\{i,j\}}$ of the concerned singularity. Therefore, it is always possible to express the residues solely in terms of gamma functions, and to express the polygamma factors as derivatives of those gamma functions. As an example, the explicit form of $R_{\{1,2\}}$ is
\begin{align}
    & R_{\{1,2\}} = - \frac{2 r_1^2}{3 \sqrt{\pi}} \sum_{m=0}^\infty \frac{\Gamma (m+2) \Gamma \left(m+\frac{5}{2}\right) \Gamma (m+4)}{\Gamma \left(m+\frac{9}{2}\right) \Gamma (m+5)} \frac{r_1^m}{m!} \nonumber \\
    & \times \bigg[ \left(-\psi ^{(0)}(m+2)+\psi ^{(0)}(m+3)-\psi ^{(0)}(m+4)+\psi ^{(0)}\left(m+\frac{9}{2}\right)-\log \left(\frac{r_1}{4r_2}\right)-\frac{5}{3}\right)^2  \nonumber \\
    & \quad +\psi ^{(1)}(m+2)-\psi ^{(1)}(m+3)+\psi ^{(1)}(m+4)-\psi ^{(1)}\left(m+\frac{9}{2}\right)+\frac{\pi ^2}{3}+\frac{31}{9} \bigg]
\end{align}

The above was obtained by applying the following Cauchy's theorem operator
\begin{align}
    \frac{1}{2} \frac{\partial^2}{\partial s^2} - \frac{\partial^2}{\partial s \partial t} + \frac{1}{2} \frac{\partial^2}{\partial t^2}
\end{align}
to
\begin{align}
    \mathcal{H}(s,t) &= - r_1^{m-s+2} r_2^{-t} \frac{\Gamma (1-s) \Gamma (s+1) \Gamma (1-t) \Gamma (2-t) \Gamma (t+1) \Gamma (m-s+2) \Gamma (m-s+4)}
    {\Gamma \left(\frac{5}{2}-t\right) \Gamma (m-s+3) \Gamma \left(m-s+\frac{9}{2}\right) \Gamma (m-s-t+1) \Gamma (m-s-t+5)} \nonumber \\
    & \times  \Gamma (1-s-t) \Gamma (s+t+1) \Gamma \left(m-s-t+\tfrac{5}{2}\right) \Gamma (m-s-t+3)
\end{align}
and thereafter setting $s=t=0$ (see \cite{Friot:2011ic} for details on the general computational procedure). $R_{\{1,2\}}$ can therefore be expressed more compactly as
\begin{align}
    \bigg[ \left( \frac{1}{2} \frac{\partial^2}{\partial \alpha^2} - \frac{\partial^2}{\partial \alpha \partial \beta} + \frac{1}{2} \frac{\partial^2}{\partial \beta^2} \right) H_{\{1,2\}} \bigg]_{\alpha=\beta=0}
\end{align}
where
\begin{multline}
    H_{\{1,2\}} =  - \frac{r_1^{2-\alpha}}{r_2^\beta} \frac{\Gamma (1-\alpha ) \Gamma (2-\alpha ) \Gamma (4-\alpha ) \Gamma (\alpha +1) \Gamma (1-\beta ) \Gamma (2-\beta ) \Gamma (\beta +1)  \Gamma \left(\tfrac{5}{2}-\alpha -\beta \right)}
    {\Gamma (3-\alpha ) \Gamma \left(\frac{9}{2}-\alpha \right) \Gamma \left(\frac{5}{2}-\beta \right)  \Gamma (5-\alpha -\beta)} \\
    \times \Gamma (3-\alpha-\beta) \Gamma (\alpha +\beta +1)
    {}_5F_4 \left[ \begin{array}{c}
	    1, 2-\alpha, 4-\alpha, \tfrac{5}{2}-\alpha-\beta, 3-\alpha-\beta \\
    	3-\alpha, \tfrac{9}{2}-\alpha, 1-\alpha-\beta, 5-\alpha-\beta \\
    \end{array}	\bigg| r_1 \right]
\end{multline}
and where we have replaced the variables $s$ and $t$ by the parameters $\alpha$ and $\beta$, and expressed the sum over $m$ in terms of the generalised hypergeometric function, ${}_5F_4$. The advantage of this notation is that it is concise, and that by expressing the single series as ${}_p F_q$, or the double series as Kamp\'e de F\'eriet series one may perform analytic continuations on these results easily if needed.

It is in some cases possible to further simplify the results and express them in terms of a single parameter. For example, $R_{\{1,2\}}$ may be expressed as
\begin{align}
    \left[ \left( \frac{2 \pi^2}{3} +  \frac{\partial^2}{\partial \alpha^2} \right) h_{\{1,2\}}  \right]_{\alpha=0}
\end{align}
where 
\begin{align}
    h_{\{1,2\}} &= -\frac{\sqrt{\pi}}{32} \frac{r_1^{\alpha+2}}{r_2^{\alpha}} \frac{\Gamma (2-\alpha ) \Gamma (\alpha +2) \Gamma (\alpha+4)}
    {\Gamma \left(\frac{5}{2}-\alpha \right) \Gamma (\alpha +3) \Gamma \left(\alpha +\frac{9}{2}\right)}
    {}_4F_3 \left[ \begin{array}{c}
	    3, 2+\alpha, 4+\alpha, \tfrac{5}{2} \\
	    5, 3+\alpha, \tfrac{9}{2}+\alpha \\
    \end{array}	\bigg| r_1 \right]
\end{align}

In the rest of this paper, we have chosen to express the results in the most compact notation possible. However, for illustrative purposes, we have given both forms (\textit{i.e} simplified and non-simplified) of the results in the electron case (see Table~\ref{Table:ElectronResults} and Table~\ref{Table:ElectronResultsLong}).

\begin{table}
\renewcommand{\arraystretch}{1.75}
\center
\begin{tabular}{ |c|c|c| } 
 \hline
 Singularity & Label & Residue \\ 
 \hline
 \hline
  
 $(-1-m,-1-n)$ & $R_{\{1,1\}}$ & $ \left[ \frac{\partial}{\partial \alpha} H_{\{1,1\}} \right]_{\alpha=0}$ \\  
 $(-2-m,0)$ & $R_{\{1,2\}}$ & $\bigg[ \left( \frac{1}{2} \frac{\partial^2}{\partial \alpha^2} - \frac{\partial^2}{\partial \alpha \partial \beta} + \frac{1}{2} \frac{\partial^2}{\partial \beta^2} \right) H_{\{1,2\}} \bigg]_{\alpha=\beta=0}$ \\ 
 $(-3-m,1)$ & $R_{\{1,3\}}$ & $ \left[ \left( \frac{\partial}{\partial \beta} -\frac{\partial}{\partial \alpha} \right) H_{\{1,3\}}  \right]_{\alpha=\beta=0}$ \\ 
 $(-4-m-n,2+m)$ & $R_{\{1,4\}}$ & $\bigg[ \left( \frac{1}{2} \frac{\partial^2}{\partial \alpha^2} - \frac{\partial^2}{\partial \alpha \partial \beta} + \frac{1}{2} \frac{\partial^2}{\partial \beta^2} \right) H_{\{1,4\}} \bigg]_{\alpha=\beta=0}$ \\
 \hline
\end{tabular}
\caption{Cone 1 results (non-simplified). Expressions for the $H_{\{1,i\}}$ can be found in Eqs.(\ref{Ap_el1ns})-(\ref{Ap_el4ns}) of the Appendix.}
\label{Table:ElectronResultsLong}
\end{table}

The $h_{\{1,i\}}$ and the $H_{\{1,i\}}$ are given in the Appendix.

\subsubsection{Convergence region: external electron \label{conv_KdF}}

The region of convergence of the corresponding series representation is straightforward to derive from the convergence properties of generalized hypergeometric and Kamp\'e de F\'eriet series \cite{Srivastava}.

For the generalized hypergeometric series 

\begin{equation}
{}_pF_q \left[ \begin{array}{c}
	    \alpha_1,..., \alpha_p \\
	    \beta_1,..., \beta_q \\
    \end{array}	\bigg| z \right]=\sum_{n=0}^\infty\frac{(\alpha_1)_n...(\alpha_p)_n}{(\beta_1)_n...(\beta_q)_n}\frac{z^n}{n!}\ ,
\end{equation}
they read (assuming that none of the parameters is zero or a negative integer):
\medskip

$(i)$ Convergence for $\vert z\vert<\infty$ if $p\leq q$,
     
$(ii)$ Convergence for $\vert z\vert<1$ if $p=q+1$,

$(iii)$ Divergence for all $z\neq 0$ if $p>q+1$.
\medskip

\noindent Furthermore, if we define 
\begin{equation}
\omega=\sum_{j=1}^q\beta_j-\sum_{j=1}^p\alpha_j\ ,
\label{omega}
\end{equation}
then the ${}_pF_q$ series, with $p=q+1$ is
\medskip
    
$(i)$ absolutely convergent for $\vert z\vert=1$ if $\text{Re}(\omega)>0$,
     
$(ii)$ conditionally convergent for $\vert z\vert=1$, $z\neq 1$ if $-1<\text{Re}(\omega)\leq0$,

$(iii)$ divergent for $\vert z\vert=1$ if $\text{Re}(\omega)\leq-1$.
\bigskip
    
For the Kamp\'e de F\'eriet double hypergeometric series (with the following notation in the l.h.s of Eq.(\ref{KdFdef}): $(a_p)\doteq(a_1,...,a_p))$
\begin{multline}
    F^{p:q;k}_{l:r;s} 
    \left[ \begin{array}{c}
	    (a_p): (b_q); (c_k) \\
	    (\alpha_l): (\beta_r); (\gamma_s)  \\
    \end{array} \Bigg| x,y \right]
    \doteq \sum_{m,n=0}^\infty
    \frac{\prod_{j=1}^p (a_j)_{m+n} \prod_{j=1}^q (b_j)_m \prod_{j=1}^k (c_j)_n }
    {\prod_{j=1}^l (\alpha_j)_{m+n} \prod_{j=1}^r (\beta_j)_m \prod_{j=1}^s (\gamma_j)_n }
    \frac{x^m}{m!} \frac{y^n}{n!}
    \label{KdFdef}
\end{multline} 
we have

$(i)$ Convergence for $\vert x\vert <\infty$ and $\vert y\vert <\infty$ if $p+q<l+m+1$ and $p+k<l+n+1$.

$(ii)$ If $p+q=l+m+1$ and $p+k<l+n+1$, the convergence is for $\vert x\vert ^{\frac{1}{(p-l)}}+\vert y\vert ^{\frac{1}{(p-l)}}<1$ if $p>l$, or for $\text{max}\{\vert x\vert ,\vert y\vert\}<1$ if $p\leq l$.

From the results above and Table \ref{Table:ElectronResults} it is easy to find that the convergence region of the r.h.s. of Eq.(\ref{electrons_results}) is simply
\begin{equation}
\mathcal{R}_e=\left\{\left|\frac{r_1}{r_2}\right|<1\hspace{0.2cm} \textrm{and}\hspace{0.2cm} |r_2|<1\right\}. \label{C2conv}
\end{equation}

This region is plotted in Figure \ref{region_conv} with the label "Cone 1".  In the case where the three different leptons are involved, $r_1=\frac{m^2_e}{m^2_\tau}$ and $r_2=\frac{m^2_e}{m^2_\mu}$ is the unique possibility to satisfy these constraints.

Note that it is in fact possible to include the boundaries in the convergence region of Eq.(\ref{C2conv}), \textit{i.e} for $\frac{r_1}{r_2}=1$ or $r_2=1$. For this one has to consider each of the two generalized hypergeometric series of Eqs.(\ref{Ap_el2}) and (\ref{Ap_el3}) in the Appendix and see that condition $(i)$ after Eq.(\ref{omega}) applies. A similar analysis can be performed on the double series of Eqs.(\ref{Ap_el1}) and (\ref{Ap_el4}) by writing each of the latter as sums of generalized hypergeometric series in either one or the other variable and check that condition $(i)$ is also satisfied.

\subsection{The muon case (Cone 2)\label{QEDmuon}}

The muon case corresponds to Cone 2 in Figure \ref{conesQED}. Therefore, the different sets of singularities to consider in the $(\textrm{Re}(s),\textrm{Re}(t))$-plane are the following, where as before $m$ and $n$ are any non-negative integers:
\begin{itemize}
\item Isolated terms: $(-1,0)$, $(-1,1)$, $\left(-1,\frac{3}{2}\right)$, $(-2,1)$.
\item Single series contributions: $(-2-m,0)$, $(-3-m,1)$, $(-3-m,2+m)$, $(-2-m,2+m)$, $(-1-m,2+m)$, $(-1-m,3+m)$.
\item Double series contributions: $(-4-m-n,2+m)$,$(-1-m,4+m+n)$.
\end{itemize}

Three of them have already been considered during the calculation of Cone 1, in the electron case. One should however keep in mind that in Cone 2, their corresponding transformation law \cite{Friot:2011ic} will not be the same as in Cone 1 so that these three sets of residues will not give the same analytic expression in Cone 1 and Cone 2.

\subsubsection{Full result}

The series representation extracted from Eq.(\ref{MBQED}) by summing the residues of Cone 2 is:
\begin{equation}
A_{3,\mu}^{(6)}(\sqrt{r_1},\sqrt{r_2})=\frac{\sqrt{\pi}}{8}\sum_{i=1}^{12} R_{\{2,i\}}\  \hspace{5mm} \text{where} \hspace{5mm} r_1 = \frac{m_\mu^2}{m_\tau^2}, \quad r_2 = \frac{m_\mu^2}{m_e^2}.
\label{muon_results}
\end{equation} 
As in the electron case, the correspondence of the singularity points and their residues $R_{\{2,i\}}$ are presented in Table~\ref{Table:MuonResults}, the explicit forms of some of the residues being relegated to the appendix for lack of space in the main body of the paper (see Eqs.(\ref{Ap_mu1})-(\ref{Ap_mu8})).

\subsubsection{Convergence region: external muon}

Using the results presented in Section \ref{conv_KdF}, it is easy to conclude that the convergence region of the r.h.s. of Eq.(\ref{muon_results}) is 
\begin{equation}
\mathcal{R_\mu}=\left\{\left|r_1\right|<1 \hspace{0.2cm} \textrm{and}\hspace{0.2cm} |r_2|>1\right\}.
\label{Rmu}
\end{equation}
See Figure \ref{region_conv} for a plot of this region (labelled "Cone 2"). As in the electron case, it is possible to include the boundaries in this convergence region. In the case where the three different leptons are involved, the only phenomenological situation which satisfies Eq.(\ref{Rmu}) is when $r_1=\frac{m^2_\mu}{m^2_\tau}$ and $r_2=\frac{m^2_\mu}{m^2_e}$.

\begin{table}
\renewcommand{\arraystretch}{1.72}
\center
\begin{tabular}{ |c|c|c| } 
\hline Singularity & Label & Residue \\ 
\hline
\hline
 $(-1,0)$ & $R_{\{2,1\}}$ &  $-\frac{16}{135 \sqrt{\pi}} r_1 \left(\frac{1}{2}-\log(r_2)\right)$ \\ 
 $(-1,1)$ & $R_{\{2,2\}}$ & $\frac{16}{15\sqrt{\pi}} \frac{r_1}{r_2}$ \\ 
 $(-2,1)$ & $R_{\{2,3\}}$ & $-\frac{8}{105 \sqrt{\pi}} \frac{r_1^2}{r_2}$ \\ 
 $(-1,3/2)$ & $R_{\{2,4\}}$ & $-\frac{32 \pi^{3/2}}{45} \frac{r_1}{r_2^{3/2}}$ \\
 $(-2-m,0)$ & $R_{\{2,5\}}$ & $ \left[ \left( \frac{1}{2} \frac{\partial^2}{\partial \alpha^2} - \frac{\partial^2}{\partial \alpha \; \partial \beta} \right) h_{\{2,5\}} \right]_{\alpha=\beta=0}$ \\
 $(-3-m,1)$ & $R_{\{2,6\}}$ & $  \left[ \frac{\partial}{\partial \alpha}  h_{\{2,6\}} \right]_{\alpha=0}$ \\ 
 $(-3-m,2+m)$ & $R_{\{2,7\}}$ & $  \left[ \frac{\partial}{\partial \alpha}  h_{\{2,7\}} \right]_{\alpha=0}$ \\  
 $(-2-m,2+m)$ & $R_{\{2,8\}}$ & $ \left[ \frac{\partial}{\partial \alpha}  h_{\{2,8\}} \right]_{\alpha=0}$ \\  
  $(-1-m,2+m)$ & $R_{\{2,9\}}$ & $  \left[ \frac{\partial^2}{\partial \alpha^2}  h_{\{2,9\}} \right]_{\alpha=0}$ \\  
  $(-1-m,3+m)$ & $R_{\{2,10\}}$ & $  \left[ \frac{\partial^2}{\partial \alpha^2}  h_{\{2,10\}} \right]_{\alpha=0}$ \\ 
  $(-4-m-n, 2+m)$ & $R_{\{2,11\}}$ & $   \left[ \left( \frac{1}{2} \frac{\partial^2}{\partial \alpha^2} -  \frac{\partial^2}{\partial \alpha \; \partial \beta} \right) h_{\{2,11\}} \right]_{\alpha=\beta=0}$ \\
 $(-1-m,4+m+n)$ & $R_{\{2,12\}}$ & $  \left[ \frac{\partial}{\partial \alpha}  h_{\{2,12\}} \right]_{\alpha=0}$ \\  
 \hline
\end{tabular}
\caption{Cone 2 results. Expressions for the $h_{\{2,i\}}$ can be found in Eqs.(\ref{Ap_mu1})-(\ref{Ap_mu8}) of the Appendix. The leading contributions are $R_{\{2,1\}}$, $R_{\{2,5\}}$ and $R_{\{2,2\}}$.}
\label{Table:MuonResults}
\end{table}

\subsection{The $\tau$ lepton case (Cone 3)\label{QEDtau}}

The $\tau$ lepton case falls in the convergence region associated to Cone 3 (see Figure \ref{conesQED}), and it is the hardest from the computational point of view for two reasons: there are a lot of different types of singularities in this cone, and here one also has to take care of the cancellation, or reduction of multiplicity, of different sets of singularities due to gamma functions in the denominator of the MB integrand (a cancellation also happened in the muon case but only for one set of singularities). 

With $m$ and $n$ any positive integer, one may as usual exhibit all sets of singularities contributing to the cone. There are 25 different sets to consider:
\begin{itemize}
\item Isolated terms: $(0,0), (0,1), (0,2), \left(-\frac{1}{2},1\right), \left(-1,\frac{3}{2}\right), \left(\frac{1}{2},0\right), \left(\frac{1}{2},1\right), (1,0), (1,1), \left(\frac{3}{2},0\right), \\ \left(\frac{3}{2},1\right), (2,0)$.
\item Single series contributions: $(0,3+m), \left(-\frac{1}{2},2+m\right), (-1-m,2+m), (-1-m,3+m), \left(\frac{1}{2},2+m\right), (1,2+m), \left(\frac{3}{2},2+m\right), (2+m,1), \left(\frac{5}{2}+m,0\right), (3+m,0)$.
\item Double series contributions: $\left(-\frac{3}{2}-m,2+m+n\right)$, $(2+m,2+n)$, $\left(\frac{5}{2}+m,2+n\right)$.
\end{itemize}
\subsubsection{Full result}

The series representation extracted from Eq.(\ref{MBQED}) by summing the residues of Cone 3 is:
\begin{align}
A_{3,\tau}^{(6)}(\sqrt{r_1},\sqrt{r_2})=\frac{\sqrt{\pi}}{8}\sum_{i=1}^{25} R_{\{3,i\}}\ \hspace{5mm} \text{where} \hspace{5mm} r_1 = \frac{m_\tau^2}{m_\mu^2}, \quad r_2 = \frac{m_\tau^2}{m_e^2}.
\label{tau_results}
\end{align} 
As in the preceding electron and muon cases, the correspondence of the singularity sets and their residues  $R_{\{3,i\}}$ are presented in Table~\ref{Table:TauResults} and the explicit forms of some of the residues is relegated to the appendix (see Eqs.(\ref{Ap_tau1})-(\ref{Ap_tau15})).

\begin{table}
\renewcommand{\arraystretch}{1.75}
\center
\begin{adjustbox}{width=0.90\textwidth}
\begin{tabular}{ |c|c|c| } 
 \hline
 Singularity & Label & Residue \\ 
 \hline
 \hline
  $(0,0)$ & $R_{\{3,1\}}$ & $\frac{8}{9 \sqrt{\pi}} \left[ \left( \frac{25}{6}-\log (r_1) \right) \left( \frac{25}{6}-\log (r_2) \right) + \frac{1}{4} + \frac{2\pi^2}{3} \right]$ \\
  $(0,1)$ & $R_{\{3,2\}}$ & $-\frac{16}{3\sqrt{\pi} } \frac{1}{r_2} \left[ \left( \frac{13}{6}-\log (r_1) \right)^2 + \frac{259}{36}+\frac{2 \pi^2}{3} \right]$ \\
  $(0,2)$ & $R_{\{3,3\}}$ & $h_{\{3,3\}}$ \\
  $(\tfrac{1}{2},0)$ & $R_{\{3,4\}}$ & $-\frac{4 \pi^{3/2}}{3r_1^{1/2}} \left[ \frac{14}{3} + \log \left( \frac{r_1}{16 r_2} \right) \right]$  \\
  $(\tfrac{1}{2},1)$ & $R_{\{3,5\}}$ & $ \frac{15\pi^{3/2}}{r_1^{1/2} r_2}$  \\
  $(1,0)$ & $R_{\{3,6\}}$ & $-\frac{32}{3 \sqrt{\pi}} \frac{1}{r_1} \left[ \left(\frac{3}{2}-\log (r_1)\right) \left(\frac{1}{2} \log \left(\frac{r_1}{r_2^2}\right)+\frac{17}{12}\right) + \frac{\pi ^2}{3}-\frac{5}{8} \right]$  \\
  $(\tfrac{3}{2},0)$ & $R_{\{3,7\}}$ & $\frac{20\pi^{3/2}}{3r_1^{3/2}} \left[ \frac{7}{3} + \log \left( \frac{r_1}{16 r_2} \right) \right]$  \\
  $(1,1)$ & $R_{\{3,8\}}$ & $-\frac{32}{\sqrt{\pi}} \frac{1}{r_1 r_2}
	\left( \frac{5}{6} +\log (r_1) \right)$  \\
  $(\tfrac{3}{2},1)$ & $R_{\{3,9\}}$ & $- \frac{35 \pi^{3/2}}{6r_1^{3/2} r_2}$ \\
  $(2,0)$ & $R_{\{3,10\}}$ & $h_{\{3,10\}}$ \\
  $(-\tfrac{1}{2},1)$ & $R_{\{3,11\}}$ & $\frac{3 \pi^{3/2} r_1^{1/2}}{2 r_2} $ \\ 
  $(-1,\tfrac{3}{2})$ & $R_{\{3,12\}}$ & $-\frac{32\pi^{3/2}r_1}{45r_2^{3/2}}$ \\
  $(0,3+m)$ & $R_{\{3,13\}}$ & $  \left[ \frac{\partial^2}{\partial \alpha \; \partial \beta} h_{\{3,13\}} \right]_{\alpha=\beta=0}$ \\
  $(-\tfrac{1}{2},2+m)$ & $R_{\{3,14\}}$ & $  \left[ \left( -\frac{5}{6} + \frac{\partial}{\partial \alpha} \right) h_{\{3,14\}} \right]_{\alpha=0}$ \\ 
  $(-1-m,2+m)$ & $R_{\{3,15\}}$ & $  \left[ \frac{\partial^2}{\partial \alpha^2} h_{\{3,15\}} \right]_{\alpha=0}$ \\
  $(-1-m,3+m)$ & $R_{\{3,16\}}$ & $  \left[ \frac{\partial^2}{\partial \alpha^2} h_{\{3,16\}} \right]_{\alpha=0}$ \\
  $(\frac{1}{2},2+m)$ & $R_{\{3,17\}}$ & $  \left[ \left( 3 + \frac{\partial}{\partial \alpha} \right) h_{\{3,17\}} \right]_{\alpha=0}$ \\
  $(1,2+m)$ & $R_{\{3,18\}}$ & $\left[ \frac{\partial}{\partial \alpha}  h_{\{3,18\}} \right]_{\alpha=0}$ \\
  $(\frac{3}{2},2+m)$ & $R_{\{3,19\}}$ & $  \left[ \left( \frac{2}{3} + \frac{\partial}{\partial \alpha} \right) h_{\{3,19\}}  \right]_{\alpha=0}$ \\
  $(2+m,1)$ & $R_{\{3,20\}}$ & $  \left[ \frac{\partial}{\partial \alpha} h_{\{3,20\}}  \right]_{\alpha=0}$ \\
  $(\tfrac{5}{2}+m,0)$ & $R_{\{3,21\}}$ & $h_{\{3,21\}}$ \\ 
 $(3+m,0)$ & $R_{\{3,22\}}$ & $  \left[ \frac{\partial^2}{\partial \alpha \; \partial \beta} h_{\{3,22\}}  \right]_{\alpha=\beta=0}$ \\ 
 $(-\tfrac{3}{2}-m,2+m+n)$ & $R_{\{3,23\}}$ & $    \left[  \frac{\partial}{\partial \alpha}  h_{\{3,23\}}  \right]_{\alpha=0}$ \\ 
 $(2+m,2+n)$ & $R_{\{3,24\}}$ & $\left[  \frac{\partial^2}{\partial \alpha \; \partial \beta}  h_{\{3,24\}}  \right]_{\alpha=\beta=0}$ \\ 
 $(\tfrac{5}{2}+m,2+n)$ & $R_{\{3,25\}}$ &  $ h_{\{3,25\}}$ \\ 
 \hline
\end{tabular}
\end{adjustbox}
\caption{Cone 3 results. Expressions for the $h_{\{3,i\}}$ can be found in Eqs.(\ref{Ap_tau1})-(\ref{Ap_tau15}) of the Appendix. The leading contributions are $R_{\{3,1\}}$, $R_{\{3,4\}}$, $R_{\{3,6\}}$, $R_{\{3,7\}}$ and $R_{\{3,10\}}$.}
\label{Table:TauResults}
\end{table}

\subsubsection{Convergence region: external tau lepton}

Once more, using the results presented in Section \ref{conv_KdF}, one concludes that the convergence region where the series representation associated to Cone 3 is valid is:
\begin{equation}
\mathcal{R}_\tau=\left\{ \left|\frac{r_1}{r_2}\right|<1\hspace{0.2cm} \textrm{and}\hspace{0.2cm} |r_1|>1 \right\}.
\label{Rtau}
\end{equation}
See Figure~\ref{region_conv} the region labelled "Cone 3". As in the two previous cases, it is in fact possible to include the boundaries in this convergence region. Here we see that with $r_1=\frac{m^2_\tau}{m^2_\mu}$ and $r_2=\frac{m^2_\tau}{m^2_e}$ the convergence constraints of Eq.(\ref{Rtau}) are satisfied.

\subsection{Other cones\label{other}} 
As already mentioned, due to the symmetry of the MB integral, it is possible to obtain the results of the other three cones by a simple interchange of $r_1$ and $r_2$ in the results that we have already obtained. One then sees in Figure~\ref{region_conv} that the whole first quadrant of the $(r_1,r_2)$-plane may be reached. Values at the boundaries of the different cones may be evaluated by using expressions of either cone. This is discussed further in Section~\ref{Sec:checks} and \ref{MBAC}.

This behaviour is therefore completely different from what can be found in the examples considered in, for instance, \cite{Friot:2011ic, Berends:1993ee, Ananthanarayan:2017qmx, Ananthanarayan:2018irl} where there were always white regions in the parameter space which could not be reached.

\section{Numerical analysis and checks\label{Sec:checks}}

Using the CODATA 2018 lepton mass ratios values \cite{CODATA2018}: $m_\mu/m_e = 206.7682830(46)$, $m_\mu/m_\tau=5.94635(40) \times 10^{-2}$, $m_e/m_\mu=4.83633169(11) \times 10^{-3}$, $m_e/m_\tau=2.87585(19) \times 10^{-4}$, $m_\tau/m_e=3477.23(23)$ and  $m_\tau/m_\mu=16.8170(11)$, we get the following values for the $A_{3,l}^{(6)}$:
\begin{align}
    & A_{3,\mu}^{(6)} = 5.27737(71) \times 10^{-4} \nonumber \\
    & A_{3,\tau}^{(6)} = 3.34778(17) \nonumber \\
    & A_{3,e}^{(6)} = 1.90972(25) \times 10^{-13}
\end{align}

We performed several internal and external checks to ensure the validity of our expressions.
The first consistency check was to compare values obtained by a high precision numerical integration of Eq.(\ref{MBQED}) and its Feynman parametrization against our analytic results. For each cone, we have tested our expressions by ensuring that there is agreement between the integral and our full analytic result, both computed numerically, to at least six orders of magnitude beyond the order of magnitude of the smallest contributing set of residues of that cone.

The second consistency check was to compare our results for Cone 2 and Cone 3 against another set of analytic expressions derived by analytically continuing the residues of Cone 1, i.e. $R_{\{1,i\}}$ for $i=1,...,4$. This results in a set of series that are numerically equivalent to $\sum R_{\{2,i\}}$ and $ \sum R_{\{3,i\}}$, but which are different in their analytic form. These were numerically evaluated, and shown to agree with the $A_{3,\mu}^{(6)}$ and $A_{3,\tau}^{(6)}$ computed directly from the $\sum R_{\{2,i\}}$ and $\sum R_{\{3,i\}}$, respectively, to at least 19 decimal places. Let us briefly describe this approach on the example of the first electron residue $R_{\{1,1\}}$ (see Table \ref{Table:ElectronResults}), which involves a Kamp\'e de F\'eriet series. The latter has the following Mellin-Barnes type integral representations \cite{KdF}

\begin{align}
	R_{\{1,1\}} = \left[ \frac{\partial}{\partial \alpha} h_{\{1,1\}} \right]_{\alpha=0}
\end{align}
where
\begin{multline}
	h_{\{1,1\}} = r_1 r_2^{1-\alpha} \Gamma (1-\alpha )^2 \Gamma (\alpha +1)^2   \int\limits_{\boldsymbol{\gamma_4} + i\mathbf{R}^2}
    \!\frac{d s}{2i\pi}\wedge\frac{d t}{2i\pi}\; (-r_1)^s (-r_2)^t \Gamma (-s)\Gamma (-t)\;
	\\
	\times \frac{ \Gamma \left(s+t+\tfrac{5}{2}-\alpha \right) \Gamma (s+t+3-\alpha)\Gamma (s+1)^2\Gamma (s+3)  \Gamma (t+1) \Gamma (t+1-\alpha) \Gamma (t+3-\alpha) }{\Gamma (s+t+1-\alpha ) \Gamma (s+t+5-\alpha) \Gamma (s+2)\Gamma \left(s+\frac{7}{2}\right) \Gamma (t+2-\alpha) \Gamma \left(t+\tfrac{7}{2}-\alpha \right)}
\label{Eq:R14AC}
\end{multline}
and $\gamma_4 \doteq(\textrm{Re}(s),\textrm{Re}(t))\in]-1,0[\times]-1,0[$ (for $\alpha<0$).

Solving this MB representation in appropriate cones yields the desired analytic continuations for that particular series. 

\begin{figure}[hbtp]
\centering
\includegraphics[height=7cm]{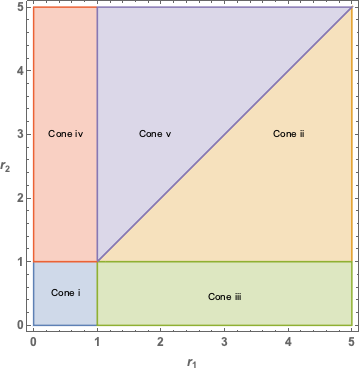}
\caption{Regions of convergence of the series representations of the MB integral of Eq.(\ref{Eq:R14AC}) and associated cones.\label{Fig:R14ACRoC}}
\end{figure}

The convergence regions of the series expansions obtainable from a direct residue calculation of Eq.(\ref{Eq:R14AC}) following the method of \cite{Friot:2011ic} are shown in Figure~\ref{Fig:R14ACRoC} and labelled by their associated cone. (Note that we use roman numerals to label and distinguish the cones of Eq.(\ref{Eq:R14AC}) from those of Eq.(\ref{MBQED})). The sum of residues of Cone i obviously reproduce the Kamp\'e de F\'eriet series of $R_{\{1,1\}}$. For the muon case, where $r_1 = m_\mu^2/m_\tau^2 \sim 10^{-3}$ and $r_2 = m_\mu^2/m_e^2 \sim 10^{4}$, the residues of Cone iv have to be calculated. For the tau case where $r_1 = m_\tau^2/m_\mu^2 \sim 10^2$ and $r_2 = m_\tau^2/m_e^2 \sim 10^{7}$, those of Cone v need to be calculated. 

By performing a similar calculation on $R_{\{1,2\}}$, $R_{\{1,3\}}$ and $R_{\{1,4\}}$, and summing the results of the appropriate cones, the results of Cone 1 (external electron legs) of Eq.(\ref{MBQED}) may be analytically continued to obtain the results of  Cone 2 (external muon legs) and Cone 3 (external tau legs) of the same integral; see \cite{Shayan:Thesis} for details and results of the complete calculation. This results in series representations that are numerically equivalent, but different in form, to those obtained from a direct calculation of Eq.(\ref{MBQED}). This process did not yield expressions that were simpler than those obtained by a direct evaluation. However, results obtained by this analytic continuation approach could well confer advantages, such as simplicity of form, in other cases than the $g-2$. 

We also checked our results externally by numerically comparing them with results from the literature.

For the muon, our expressions evaluated with the CODATA 2010 \cite{Mohr:2012tt} mass ratios yield the same value of $A_{3,\mu}^{(6)}$ as given in Eq.(9) of \cite{Aoyama:2012wk}. Similarly, we get agreement with Eq.(22) of \cite{Laporta:1993ju} evaluated using the values of the 1992 PDG \cite{Hikasa:1992je}, and with Eq.(16) of \cite{Passera:2004bj} evaluated using CODATA 2002 \cite{Mohr:2005zz} inputs, the latter of which is based on the first few terms of the asymptotic and eikonal expansion derived expressions of \cite{Czarnecki:1998rc}. Our expression differs numerically from the value given in Eq.(33) of \cite{Samuel:1990qf} at the sixth decimal place. The same integral is evaluated in \cite{Passera:2006gc, Eidelman:2007sb} with CODATA 2006 \cite{Mohr:2008fa} mass ratios to yield $A_{3,\mu}^{(6)}$, values with which our expressions are in agreement. Our results for $A_{3,\tau}^{(6)}$, however, yield a numerical value that differs from the older literature values \cite{Narison:2001jt, Samuel:1990su} but agrees with \cite{Passera:2006gc}, as already mentioned in the introduction and in Section \ref{Sec:QED}. For the electron, our $A_{3,e}^{(6)}$ expression agrees numerically with the value given in \cite{Passera:2006gc}, based on the integral of \cite{Samuel:1990qf} and 2004 CODATA values, as well as with the result given in Eq.(12) of \cite{Kurz:2013exa} that uses the PDG 2012 values as inputs. And finally, within the latter's uncertainty range, we obtain the same result as \cite{Aoyama:2012wj} using the CODATA 2010 \cite{Mohr:2012tt} mass ratios.

Note that the diagram of Figure~\ref{g-2_3loop} appears in the expansion of $a_l$ not only as $A_{3,l}^{(6)}$, but also as one of the several diagrams that constitute $A_{2,l}^{(6)}(m_l/m_{l'})$ and $A_{1,l}^{(6)}$.  Table~\ref{Table:TwoMassCones} gives, among others, the various mass configurations possible for Figure~\ref{g-2_3loop} with two masses, and the cones corresponding to their evaluation. As all these cases lie on the boundary of the convergence regions of two cones (in passing, the mass independent case can be computed with any of the 6 different series representations) the convergence of the series can be slow. However, by comparing the literature values of these diagrams with our expressions, we obtain another check of the latter. Note that for the cases where there are identical leptons in the loops, an overall factor of $1/2$ is required in front of Eq.~\ref{MBQED}.

By setting all masses equal, we obtain the scaleless diagram $A_{1,l}^{(6)}$, which corresponds to the `triple point' $(1.0,1.0)$ in Figure~\ref{region_conv}. Despite the slow convergence of our series at this point, we find that as the number of terms in the sums is increased, the results of all cones numerically tend to the value given by Eq.(8) or (9) of \cite{Laporta:1993ju} or more explicitly by Eq.(4.43) of \cite{Aguilar:2008qj}: $\frac{8}{3} \zeta(3)-\frac{4\pi^2}{135}-\frac{943}{324}$.
The same equations of the paper \cite{Laporta:1993ju} give the closed form result for the cases where both leptons in the internal loops are identical, but different from the one on the external legs, and where only one of the leptons in the internal loops is the same as the one on the external legs. And Eqs.(12), (13), (16) and (17) of \cite{Laporta:1993ju} give the expansions in the mass ratio of Eqs.(8) and (9). For the case where the muon is on the external legs, with two internal electron loops, we get from our expression a fast converging numerical agreement with the value calculated using Eq.(8) of \cite{Laporta:1993ju}. With two internal tau loops the agreement is also very good, but we use Eq.(16) of \cite{Laporta:1993ju} to compare because imaginary contributions appear when using Eq.(8). With only one internal electron or tau loop, we get values that are converging in a slower way to those obtained using Eq.(9) of the aforementioned paper as we increase the number of terms in the sum. The same is true in case of tau external legs, where we get very good and fast converging numerical agreement with Eq.(8) or (9) of \cite{Laporta:1993ju} for loops with two electrons or two muons, and slower convergence towards the literature values otherwise. For the case with electron masses in the external legs, we do not use the closed form expressions of Eq.(8) and (9) of \cite{Laporta:1993ju}, but rather the expansion in the mass ratios given in Eqs.(16)-(17) of the same paper. As with the other cases, we get very good agreement when the two fermionic loops carry the same mass, and slower convergence otherwise. All these "two masses" results are also in agreement with the numerical integration of the Mellin-Barnes representations given in Eqs.(4.23) and (4.25) of \cite{Aguilar:2008qj}.

\section{Convergence and analytic continuation properties of Kamp\'e de F\'eriet series derived from Mellin-Barnes integrals\label{MBAC}}

The analysis of the previous sections showed that, for all possible numerical values of the mass ratios $r_1$ and $r_2$, at least one of the six different series representations derived from the MB representation can be used. As already briefly mentioned, this fact is unusual since in general Feynman diagrams having $N$-fold MB  representations with $N>1$, depending on some parameters $r_1$, ..., $r_N$, result in cones and series representations whose associated convergence regions do not collectively cover the entire possible $(r_1,...,r_N)$-space: some ("white") regions  exist such that none of the series representations extracted from a standard residues computation of the MB integral will converge (nor be diverging asymptotic expansions). We recall that a few examples of such integrals can be found for instance in \cite{Friot:2011ic, Berends:1993ee, Ananthanarayan:2017qmx, Ananthanarayan:2018irl} and that for this kind of integrals, it is necessary to analytically continue the results of a particular cone with an alternative method in order to derive results valid in the unreachable white region, as done for instance in \cite{Ananthanarayan:2019icl}.

The white regions do not appear for Feynman diagrams whose MB representation is 1-fold, because in this case the corresponding series are made of generalized hypergeometric functions whose analytic continuations can be trivially derived from the MB integral, and cover the entire parameter space (except possibly when the absolute value of the parameter is equal to unity).

In our present case of study, where $N=2$, it is then clear that the lack of white regions is related to the particular properties of the double series that appear in the series representations derived in the previous sections. A quick look in the Appendix shows that there are two different types of double series in all the mathematical expressions derived from the different cones, namely the Kamp\'e de F\'eriet series
\begin{equation}
 F^{2:3;3}_{2:2;2} \left[ \begin{array}{c}
    	a_1, a_2: b_1, b_2, b_3; c_1, c_2, c_3 \\
        \alpha_1, \alpha_2: \beta_1, \beta_2; \gamma_1, \gamma_2
\end{array}	\bigg| x, y \right]
\label{F23,22}
\end{equation}
and
\begin{equation}
 F^{2:3;2}_{2:2;1} \left[ \begin{array}{c}
    	a_1, a_2: b_1, b_2, b_3; c_1, c_2 \\
        \alpha_1, \alpha_2: \beta_1, \beta_2; \gamma_1\end{array}	\bigg| x, y \right]\ .
\label{F232,221}
\end{equation}
These Kamp\'e de F\'eriet double series also appear in our formulas with different arguments, but one can perform the analysis of this section to these cases also and the conclusions will be the same. Note that Eq.(\ref{F232,221}) can in fact trivially be transformed to the form of Eq.(\ref{F23,22}) by including a $\tfrac{(c_3)_n}{(c_3)_n}$ factor in the sum over $n$ of the definition of Eq.(\ref{F232,221}) given in Eq.(\ref{KdFdef}). Therefore, let us focus on Eq.(\ref{F23,22}).

From Section \ref{conv_KdF}, one can deduce that the convergence region of $F^{2:3;3}_{2:2;2}$ is the simple region $\vert x\vert<1\wedge\vert y\vert<1$. 
And in Section \ref{Sec:checks} we have seen that from the MB representation of $F^{2:3;3}_{2:2;2}$ (given in Eq.(\ref{Eq:R14AC}) for specific values of its coefficients, which have no consequence on the convergence properties), one can show that the convergence regions associated to each of the possible series representations derived from the MB integral are those of Figure \ref{Fig:R14ACRoC}, where there is no white region.

\begin{table}
\begin{tabular}{ |c|c|c| } 
 \hline
 Cone 1 & Cone 2 & Cone 3  \\ 
 \hline
 \hline
 $(m,n)$  & $(-a_j-m-n,m)$, $j\in[1,1+i]$ & $(-a_j-m-n,m)$, $j\in[1,1+i]$  \\
  &  $(-b_j-m,n)$, $j\in[1,1+k]$ & $(-b_j-m,b_j-a_p+m-n), \begin{cases}
    j\in[1,1+k]\\
    p\in[1,1+i]
  \end{cases}$   \\
&   & $(-b_j-m,-c_p-n), \begin{cases}
    j\in[1,1+k]\\
    p\in[1,1+l]
  \end{cases}$   \\
\hline
\end{tabular} \hspace{7mm}
\begin{tabular}{ |c|c| } 
 \hline
 Cone 4 & Cone 5   \\ 
 \hline
 \hline
  $(m,-a_j-m-n)$, $j\in[1,1+i]$ & $(m,-a_j-m-n)$, $j\in[1,1+i]$   \\
    $(c_j-a_p+m-n,-c_j-m)$, $\begin{cases}
    j\in[1,1+l]\\
    p\in[1,1+i]
  \end{cases}$   & $(m,-c_j-n)$, $j\in[1,1+l]$ \\
  $(-b_j-m,-c_p-n)$, $\begin{cases}
    j\in[1,1+k]\\
    p\in[1,1+l]
  \end{cases}$ &  \\
\hline
\end{tabular}
\caption{The different sets of singularities of the 5 cones of Eq.(\ref{MBKdFikl}).\label{table_general_case}}
\end{table}

In fact, the convergence and analytic continuation properties of the Kamp\'e de F\'eriet $F^{2:3;3}_{2:2;2}$ double hypergeometric series are exactly the same as those of the Appell $F_1$ double hypergeometric series. It is easy to show this from the MB representation of the latter. The similarity between the Appell $F_1$ and the Kamp\'e de F\'eriet $F^{2:3;3}_{2:2;2}$ series should not come as a complete surprise because Kamp\'e de F\'eriet series are generalisations of Appell series in the same way as generalised ${}_pF_q$ hypergeometric series are generalisations of the Gauss ${}_2F_1$ hypergeometric series. Obviously, in the latter case there is only one function to generalise, whereas in the Appell case, there are four different ones which do not satisfy the same convergence and analytic continuation properties. In particular, $F_1$ is the only Appell series whose MB integral representation gives birth to some series representations, analytic continuations of one another,  whose convergence regions are able to collectively cover the whole $(\vert x\vert,\vert y\vert)$-space (except possibly on some boundary lines at which the convergence depends on the values of the coefficients). Indeed, for the other Appell $F_2$, $F_3$ and $F_4$ functions, there are white regions in each of the corresponding convergence regions plots of their series representations derived from the usual residue evaluation of their MB representations. 

From the similarities between the representation of $F_1$ as a Kamp\'e de F\'eriet series
\begin{equation}
 F_1(a,b,c;\alpha;x,y)=F^{1:1;1}_{1:0;0} \left[ \begin{array}{c}
    	a: b; c \\
        \alpha:-;-
\end{array}	\bigg| x, y \right]
\label{F11,10}
\end{equation}

and Eqs.(\ref{F23,22}) and (\ref{F232,221}), it is natural to conjecture that the more general Kamp\'e de F\'eriet series of the type 
\begin{equation}
 F^{1+i: 1+k; 1+l}_{1+i: k; l} \left[ \begin{array}{c}
    	a_1,..., a_{1+i}: b_1,...,b_{1+k}; c_1, ..., c_{1+l} \\
        \alpha_1,...,\alpha_{1+i}: \beta_1,..., \beta_{k}; \gamma_1, ...,\gamma_l
\end{array}	\bigg| x, y \right]
\label{KdFikl}
\end{equation}
do not have white regions in their convergence picture, which we now prove. 

The Mellin-Barnes representation of the Kamp\'e de F\'eriet function shown in Eq.(\ref{KdFikl}) can be easily derived from  \cite{KdF} and reads
\begin{multline}
F^{1+i: 1+k; 1+l}_{1+i: k; l} \left[ \begin{array}{c}
    	(a_{1+i}): (b_{1+k}); (c_{1+l}) \\
        (\alpha_{1+i}): (\beta_{k}); (\gamma_l)
\end{array}	\bigg| x, y \right]=	\frac{\prod_{j=1}^{1+i} \Gamma (\alpha_j) \prod_{j=1}^{k} \Gamma (\beta_j) \prod_{j=1}^{l} \Gamma (\gamma_j)}
	{\prod_{j=1}^{1+i} \Gamma (a_j) \prod_{j=1}^{1+k} \Gamma (b_j) \prod_{j=1}^{1+l} \Gamma (c_j)} \\
	\times \int\limits_{\boldsymbol{\gamma} + i\mathbf{R}^2}
    \!\!\frac{d s}{2i\pi}\wedge\frac{d t}{2i\pi}\; (-x)^s (-y)^t \Gamma (-s) \Gamma (-t) 
	\frac{\prod_{j=1}^{1+i} \Gamma (a_j+s+t) \prod_{j=1}^{1+k} \Gamma (b_j+s) \prod_{j=1}^{1+l} \Gamma (c_j+t)}
	{\prod_{j=1}^{1+i} \Gamma (\alpha_j+s+t) \prod_{j=1}^{k} \Gamma (\beta_j+s) \prod_{j=1}^{l} \Gamma (\gamma_j+t)} \ ,
\label{MBKdFikl}
\end{multline}
where we recall that $(a_{1+i})\doteq (a_1,..., a_{1+i})$. 

To simplify our presentation, we suppose that the $a_j$, $b_j$ and $c_j$ are non-zero positive numbers. This allows us to deal with straight contours and $\boldsymbol{\gamma} \doteq(\textrm{Re}(s),\textrm{Re}(t))$ therefore belongs to the non empty fundamental polygon defined by the positivity constraint of each of the real parts of the gamma functions in the integrand. We also suppose that the singular structure of the integrand has no poles of multiplicity greater than 1, and that the values of the $\alpha_j$, $\beta_j$ and $\gamma_j$ cannot make the gamma functions of the denominator interfere with those of the numerator. 

Following the method of \cite{Friot:2011ic} it can be shown that this MB integral has five different cones and therefore five different double series representations. We list the sets of singularities for each cone in Table \ref{table_general_case} ($m$ and $n$ can be any positive integer or zero).

In the $g-2$ case, the MB integral of Eq.(\ref{MBQED}) is symmetric under the exchange of $r_1$ and  $r_2$, and therefore of the total number of series, three of them could be trivially obtained from the three others using that symmetry. Here, one can still use this symmetry, although since $k\neq l$ and $b_j\neq c_j$ generally in Eq.(\ref{MBKdFikl}) one has to exchange, in addition to $r_1$ and $r_2$, the values of the coefficients $b_j$ and $c_j$ as well as $k$ and $l$. Since Cone 1 obviously gives the Kamp\'e de F\'eriet series in Eq.(\ref{KdFikl}), it is sufficient to compute the series representations associated with Cone 2 and Cone 3, the two others being deduced from them.  

Let us begin with Cone 2 which has only two different types of contributions (see Table \ref{table_general_case}). In this case, one obtains 
\begin{multline}
F^{1+i: 1+k; 1+l}_{1+i: k; l} \left[ \begin{array}{c}
    	(a_{1+i}): (b_{1+k}); (c_{1+l}) \\
        (\alpha_{1+i}): (\beta_{k}); (\gamma_l)
\end{array}	\bigg| x, y \right]=	\frac{\prod_{r=1}^{1+i} \Gamma (\alpha_r) \prod_{r=1}^{k} \Gamma (\beta_r) }
	{\prod_{r=1}^{1+i} \Gamma (a_r) \prod_{r=1}^{1+k} \Gamma (b_r) }\Bigg(\\
	\times\sum_{j=1}^{1+i}\Gamma(a_j) \frac{\prod_{r=1}^{1+i,*} \Gamma (a_r-a_j) \prod_{r=1}^{1+k} \Gamma (b_r-a_j) }
	{\prod_{r=1}^{1+i} \Gamma (\alpha_r-a_j) \prod_{r=1}^{k} \Gamma (\beta_r-a_j) }\\
	\times (-x)^{-a_j} F^{1+k: 1+l; 1+i}_{1+k: l; i} \left[ \begin{array}{c}
    	a_j,((1-\beta+a_j)_k): (c_{1+l}); ((1-\alpha+a_j)_{1+i}) \\
        ((1-b+a_j)_{1+k}): (\gamma_{l}); ((1-a+a_j)_{1+i})^*
\end{array}	\bigg| \frac{y}{x}, \frac{1}{x} \right]\\
+\sum_{j=1}^{1+k}\Gamma(b_j)\frac{\prod_{r=1}^{1+k,*} \Gamma (b_r-b_j) \prod_{r=1}^{1+i} \Gamma (a_r-b_j) }
	{\prod_{r=1}^{k} \Gamma (\beta_r-b_j) \prod_{r=1}^{1+i} \Gamma (\alpha_r-b_j) }\\
	\times(-x)^{-b_j}\sum_{m=0}^\infty \sum_{n=0}^\infty\frac{1}{m!n!}\left(\frac{1}{x}\right)^my^n\frac{\prod_{r=1}^{1+i}  (a_r-b_j)_{n-m} (b_j)_m \prod_{r=1}^{k} (1-\beta_r+b_j)_m \prod_{r=1}^{1+l} (c_r)_n}
	{\prod_{r=1}^{1+i} (\alpha_r-b_j)_{n-m} \prod_{r=1}^{1+k,*}  (1-b_r+b_j)_m \prod_{r=1}^{l} (\gamma_r)_n}\Bigg)\ ,
\label{KdFiklCone2}
\end{multline}
where the $*$ superscript means that, for instance, the $r=j$ case is not considered in the product $\prod_{r=1}^{1+i,*} \Gamma (a_r-a_j)$, and that $((1-a+a_j)_{1+i})^*=(1-a_1+a_j, 1-a_2+a_j,..., 1-a_{j-1}+a_j, 1-a_{j+1}+a_j,..., 1-a_{1+i}+a_j)$.

It is easy to see, in the analytic continuation formula Eq.(\ref{KdFiklCone2}), that the first series, being a Kamp\'e de F\'eriet double hypergeometric series, converges in the region $\left\vert\frac{y}{x}\right\vert<1 \wedge \left\vert\frac{1}{x}\right\vert<1$.

The second series in Eq.(\ref{KdFiklCone2}) is not a Kamp\'e de F\'eriet series, but by the cancellation of parameters method \cite{Srivastava}, one finds that it converges as a Horn $G_2$ double hypergeometric series, \textit{i.e} in the region $\left\vert\frac{1}{x}\right\vert<1 \wedge \vert y\vert<1$, so that the convergence region of the analytic continuation of $F^{1+i: 1+k; 1+l}_{1+i: k; l}$ associated to Cone 2 and given in Eq.(\ref{KdFiklCone2}) converges for $\left\vert\frac{1}{x}\right\vert<1 \wedge \vert y\vert<1$. This region corresponds to the green  region plotted in Figure \ref{Fig:R14ACRoC}.

One can now proceed to the presentation of the results associated with Cone 3. In this case, one gets
\begin{multline}
F^{1+i: 1+k; 1+l}_{1+i: k; l} \left[ \begin{array}{c}
    	(a_{1+i}): (b_{1+k}); (c_{1+l}) \\
        (\alpha_{1+i}): (\beta_{k}); (\gamma_l)
\end{array}	\bigg| x, y \right]=	\frac{\prod_{r=1}^{1+i} \Gamma (\alpha_r) \prod_{r=1}^{k} \Gamma (\beta_r) }
	{\prod_{r=1}^{1+i} \Gamma (a_r) \prod_{r=1}^{1+k} \Gamma (b_r) }\Bigg(\\
	\sum_{j=1}^{1+i}\Gamma(a_j) \frac{\prod_{r=1}^{1+i,*} \Gamma (a_r-a_j) \prod_{r=1}^{1+k} \Gamma (b_r-a_j) }
	{\prod_{r=1}^{1+i} \Gamma (\alpha_r-a_j) \prod_{r=1}^{k} \Gamma (\beta_r-a_j) }\\
	\times (-x)^{-a_j} F^{1+k: 1+l; 1+i}_{1+k: l; i} \left[ \begin{array}{c}
    	a_j,((1-\beta+a_j)_k): (c_{1+l}); ((1-\alpha+a_j)_{1+i}) \\
        ((1-b+a_j)_{1+k}): (\gamma_{l}); ((1-a+a_j)_{1+i})^*
\end{array}	\bigg| \frac{y}{x}, \frac{1}{x} \right]\\
+\frac{\prod_{r=1}^{l} \Gamma (\gamma_r) }
	{\prod_{r=1}^{1+l} \Gamma (c_r)  }\sum_{j=1}^{1+k}\left(\sum_{p=1}^{1+l}\Gamma (b_j)\Gamma (c_p)\frac{\prod_{r=1}^{1+i} \Gamma (a_r-b_j-c_p)\prod_{r=1}^{1+k,*} \Gamma (b_r-b_j) \prod_{r=1}^{1+l,*} \Gamma (c_r-c_p)  }
	{\prod_{r=1}^{1+i} \Gamma (\alpha_r-b_j-c_p)\prod_{r=1}^{k} \Gamma (\beta_r-b_j) \prod_{r=1}^{l} \Gamma (\gamma_r-c_p)   }\right.\\
	\times(-x)^{-b_j}(-y)^{-c_p}F^{1+i: 1+k; 1+l}_{1+i: k; l} \left[ \begin{array}{c}
    	((1-\alpha+b_j+c_p)_{1+i}): b_j,((1-\beta+b_j)_k); c_p,((1-\gamma+c_p)_l) \\
        ((1-a+b_j+c_p)_{1+i}): ((1-b+b_j)_{1+k})^*; ((1-c+c_p)_{1+l})^*
\end{array}	\bigg| \frac{1}{x}, \frac{1}{y} \right]\\
+\sum_{p=1}^{1+i}\Gamma (b_j)\Gamma (-b_j+a_p)\frac{\prod_{r=1}^{1+i,*} \Gamma (a_r-a_p)\prod_{r=1}^{1+k,*} \Gamma (b_r-b_j) \prod_{r=1}^{1+l} \Gamma (c_r+b_j-a_p)  }
	{\prod_{r=1}^{1+i} \Gamma (\alpha_r-a_p)\prod_{r=1}^{k} \Gamma (\beta_r-b_j) \prod_{r=1}^{l} \Gamma (\gamma_r+b_j-a_p)   }\\
	\times(-x)^{-b_j}(-y)^{b_j-a_p}\sum_{m=0}^\infty \sum_{n=0}^\infty\frac{1}{m!n!}\left(\frac{y}{x}\right)^m\left(\frac{1}{y}\right)^n\frac{\prod_{r=1}^{1+l}  (c_r+b_j-a_p)_{m-n} }
	{(1-b_j+a_p)_{m-n}\prod_{r=1}^{l} (\gamma_r+b_j-a_p)_{m-n} }\\
	\left.\left. \frac{(b_j)_m\prod_{r=1}^{k}  (1-\beta_r+b_j)_m  \prod_{r=1}^{1+i} (1-\alpha_r+a_p)_n }
	{\prod_{r=1}^{1+k,*} (1-b_r+b_j)_m \prod_{r=1}^{1+i,*} (1-a_r+a_p)_n }\right)\right)\\
\label{KdFiklCone3}
\end{multline}

A convergence study similar to the one performed above shows that one finally obtains, for the series representation presented in Eq.(\ref{KdFiklCone3}), the same convergence region as the one plotted in orange in Figure \ref{Fig:R14ACRoC}, \textit{i.e} the region $\left\vert\frac{y}{x}\right\vert<1 \wedge \left\vert \frac{1}{x}\right\vert<1 \wedge \left\vert \frac{1}{y}\right\vert<1$.

It is straightforward to show that the series associated with Cone 4 and Cone 5 will converge, respectively, in the purple and in pink regions of Figure~\ref{Fig:R14ACRoC}, and one can check that our conjecture that there are no white regions for the class of Kamp\'e de F\'eriet series considered in Eq.(\ref{KdFikl}) is correct.
Therefore, we see that when dealing with this class of series, one does not need to perform non trivial analytic continuations in white regions, such as is needed for those of \cite{Ananthanarayan:2019icl} (see also \cite{AFGH17}).

We conclude this section by noting that a similar analysis can be performed in order to extract other classes of Kamp\'e de F\'eriet series that have common convergence and analytic continuation properties to the three other Appell series (and therefore also the same white regions). We do not perform such an analysis here but instead give two examples of Feynman diagrams taken from some of our recent work where there are similarities between the Kamp\'e de F\'eriet series found in their analytic expressions and the Appell $F_3$ and $F_4$ series. 
The two-loop box diagonal calculation presented in \cite{ABFP20} involves a Kamp\'e de F\'eriet $F^{2:2;2}_{3:0;0}$ double series which has the same white region as the Appell $F_3$ series. Let us recall that
  \begin{equation}
 F_3(a,b,c,d;\alpha;x,y)=F^{0:2;2}_{1:0;0} \left[ \begin{array}{c}
    	-: a,b ; c,d \\
        \alpha:-;-
\end{array}	\bigg| x, y \right]\ .
\label{F02,10}
\end{equation}

The example related to the Appell $F_4$ series comes from the chiral perturbation theory sunsets studied in \cite{Ananthanarayan:2017qmx}. In the $\bar H^\pi_{KK\eta}$ pion and $\bar H^\eta_{KK\pi}$ eta sunsets analytic expressions, one can find the Kamp\'e de F\'eriet $F^{3:1;1}_{1:2;2}$ double hypergeometric series, which has the same white region as
 \begin{equation}
 F_4(a,b;\alpha,\beta;x,y)=F^{2:0;0}_{0:1;1} \left[ \begin{array}{c}
    	a, b: -; - \\
        -:\alpha;\beta
\end{array}	\bigg| x, y \right].
\label{F4}
\end{equation}

These two examples related to $F_3$ and $F_4$ need analytic continuation procedures alternative to the traditional MB representation in order to be analytically computed using convergent series representations in their white regions. 
 
The discussion of the present section draws attention to the fact that the "Appell $F_1$" class of Kamp\'e de F\'eriet double hypergeometric series presented in Eq.(\ref{KdFikl}) has a particularly simple analytic continuation behaviour and it suggests that these nice properties are also very probably satisfied by extensions of Kamp\'e de F\'eriet series. Such a higher order class of Kamp\'e de F\'eriet multiple series would then include the $F_D^{(n)}$ Lauricella series as the simplest series of its set.

\section{Conclusion and Discussion\label{concl}}

The aim of this work was two-fold: to present complete analytic results for the three loop contributions to leptonic g-2 with two internal loops, and to use these calculations to further our understanding of the convergence and analytic continuation properties of multiple hypergeometric series that can appear in quantum field theory calculations.

In the first part of this work, we have calculated and presented complete analytic results for the three-loop QED contributions to the $g-2$ of all charged leptons with two internal loops, i.e. for all three possibilities of external legs in the Feynman diagram of Figure~\ref{g-2_3loop}, denoted in the literature by $A^{(6)}_{3,l}$ for $l=e,\mu,\tau$. In the muon case, this was the last missing piece in the puzzle of the exact results at three loop level \cite{Jegerlehner:2017gek}. Furthermore, to our knowledge, in the electron and $\tau$ lepton cases, analytic results were unknown but for the leading term in a double expansion in the mass ratios for the $\tau$ lepton case given in \cite{Narison:1977jc, Narison:2001jt}. Therefore, this work presents the first complete and exact analytic result for $A^{(6)}_{3,l}$ for all $l=e,\mu,\tau$. 

We have performed several checks of the expressions given in this paper. These included numerical comparisons with values from the literature, as well as consistency checks. One such consistency check involved the calculation of the diagram of Fig.~\ref{g-2_3loop} with two or three of the leptons being identical, i.e. the contributions of Fig.~\ref{g-2_3loop} to $A^{(6)}_{1,l}$ and $A^{(6)}_{2,l}$. The use of only two distinct masses, or one single mass, in expressions consisting of two mass ratios lands us on the boundaries of the convergence regions, which may be reached by two or more distinct series. Finding numerical agreement for expressions that were different in form was one consistency check (note that for these cases we have also found agreement with numerical results derived from the analytic expressions given in \cite{Laporta:1993ju} and from the numerical integration of the Mellin-Barnes representations of \cite{Aguilar:2008qj}). The same principle was the basis for our other self-consistency check. In this check, we have calculated a second set of expressions by analytic continuation of the results obtained in the electron case (see \cite{Shayan:Thesis} for a complete listing of these expressions), and then compared them to Eq.~(\ref{muon_results}) and Eq.~(\ref{tau_results}), which are numerically equivalent but have a different form.

Our tool for performing the calculation was the Mellin-Barnes representation, which produced results in the form of a linear combination of isolated terms and infinite single and double series consisting of products of gamma and polygamma functions, that can then be expressed in terms of generalized hypergeometric and Kamp\'e de F\'eriet series and their derivatives. These series in turn become the objects of study of the second part of this paper.

The second part of our work involved studying the analytic continuation of a class of Kamp\'e de F\'eriet series. Indeed, the generalized hypergeometric and Kamp\'e de F\'eriet series converge in a range of values of their variables. By analytic continuation, one is able to extend this range. But for the double hypergeometric series, when this analytic continuation is performed by means of the series derived from standard residues computations of MB representation, there usually exists a range of values (the white region) for which it is still not possible to derive some converging series. However, for the series appearing in the $g-2$ calculation of this paper, we find no white regions.

Inspired by this unusual analytic continuation property, in this paper we prove that for a class of Kamp\'e de F\'eriet series, namely those of type $F^{1+i:1+k;1+l}_{1+i:k;l}$, no white region can appear, and that it is possible to analytically continue these series using their MB representation to obtain convergent series for all values of their variables (except possibly, depending on the values of the parameters, on the boundaries of the convergence regions). The convergence and analytic continuation properties of the $F^{1+i:1+k;1+l}_{1+i:k;l}$ series parallel those of the Appell $F_1$ series. We also give examples of some Feynman diagrams that indicate that other classes of the Kamp\'e de F\'eriet series, whose convergence and analytic continuation properties mimic those of the other Appell series, may be found in physical situations.

In this work, therefore, we present complete analytic results at the three-loop level for the $A_{3,l}^{(6)}$ contribution to the important physical quantity $g-2$, and in the process of the calculation extend our understanding of the convergence and analytic continuation properties of the  Kamp\'e de F\'eriet series. Further investigating these properties for other classes of Kamp\'e de F\'eriet series and multiple series of higher order, of which relatively little is known, is an important direction for future research.

\vspace{1cm}

\noindent {\bf Acknowledgements}

\vspace{0.5cm}

We warmly thank David Greynat for his help in the earlier stages of this work, and for useful discussions. B.A. acknowledges partial support from the MSIL Chair of the Division of Physical and Mathematical Sciences, Indian Institute of Science, Bangalore during the course of this work. S.G. thanks Ulf-G. Meissner for supporting the research through grants. S.F. thanks the Centre for High Energy Physics, Indian Institute of Science, Bangalore, and B.A and S.G. thank the Institut de Physique Nucl\'eaire d'Orsay, Universit\'e Paris-Sud for their hospitality during the course of this work.

\newpage

\appendix

\section*{Appendix: $h_{\{i,j\}}$ and $H_{\{i,j\}}$ expressions of Tables \ref{Table:ElectronResults}, \ref{Table:ElectronResultsLong}, \ref{Table:MuonResults} and \ref{Table:TauResults}\label{SecQEDResults}}

\bigskip

\subsection*{Electron case (simplified results, see Table \ref{Table:ElectronResults}) }

\setcounter{equation}{0}
\renewcommand{\theequation}{A-\arabic{equation}}

\begin{align}
    h_{\{1,1\}} &= \frac{16}{15\sqrt{\pi}} \frac{r_1}{r_2^{\alpha-1}} \frac{ \Gamma \left(\frac{5}{2}-\alpha \right) \Gamma (3-\alpha )^2 }
    {\Gamma (2-\alpha ) \Gamma \left(\frac{7}{2}-\alpha \right) \Gamma (5-\alpha )} \nonumber\\ 
    &\hspace{2cm}\times 
    F^{2:3;3}_{2:2;2} \left[ \begin{array}{c}
    	\tfrac{5}{2}-\alpha, 3-\alpha : 1, 1, 3; 1, 1-\alpha, 3-\alpha \\
        1-\alpha, 5-\alpha : 2, \tfrac{7}{2}; 2-\alpha, \tfrac{7}{2}-\alpha
\end{array}	\bigg| r_1, r_2 \right]
\label{Ap_el1}\\
    h_{\{1,2\}} &= -\frac{\sqrt{\pi}}{32} \frac{r_1^{\alpha+2}}{r_2^{\alpha}} \frac{\Gamma (2-\alpha ) \Gamma (\alpha +2) \Gamma (\alpha+4)}
    {\Gamma \left(\frac{5}{2}-\alpha \right) \Gamma (\alpha +3) \Gamma \left(\alpha +\frac{9}{2}\right)}
    {}_4F_3 \left[ \begin{array}{c}
	    3, 2+\alpha, 4+\alpha, \tfrac{5}{2} \\
	    5, 3+\alpha, \tfrac{9}{2}+\alpha \\
    \end{array}	\bigg| r_1 \right]
    \label{Ap_el2}\\
    h_{\{1,3\}} &= - \frac{\sqrt{\pi}}{16}
    \frac{r_1^{\alpha +3}}{r_2^{\alpha+1}}
    \frac{\Gamma (\alpha +3) \Gamma (\alpha +5)}
    {\Gamma \left(\frac{3}{2}-\alpha \right) \Gamma (\alpha +2) \Gamma (\alpha +4) \Gamma \left(\alpha +\frac{11}{2}\right) }
    {}_4F_3 \left[ \begin{array}{c}
	    3, 3+\alpha, 5+\alpha, \tfrac{5}{2} \\
	    5, 4+\alpha, \tfrac{11}{2}+\alpha \\
    \end{array}	\bigg| r_1 \right]
    \label{Ap_el3}\\
    h_{\{1,4\}} &= \frac{1}{32 \sqrt{\pi}} \frac{r_1^{\alpha+4}}{r_2^{\alpha+2}} \frac{\Gamma \left(\alpha +\frac{1}{2}\right) \Gamma (\alpha +2) \Gamma (\alpha +4) \Gamma (\alpha +6)}{\Gamma (\alpha +1) \Gamma (\alpha +3) \Gamma (\alpha +5) \Gamma \left(\alpha +\frac{13}{2}\right)} \nonumber\\ 
    &\hspace{2cm}\times 
    F^{2:3;2}_{2:2;1} \left[ \begin{array}{c}
	    4+\alpha, 6+\alpha: 1, \tfrac{1}{2}+\alpha, 2+\alpha; \tfrac{5}{2}, 3 \\
        5+\alpha, \tfrac{13}{2}+\alpha : 1+\alpha, 3+\alpha; 5
    \end{array}	\bigg| \frac{r_1}{r_2}, r_1 \right]
    \label{Ap_el4}
\end{align}

\subsection*{Electron case (non-simplified results, see Table \ref{Table:ElectronResultsLong})}

\begin{align}
    H_{\{1,1\}} &= h_{\{1,1\}}
    \label{Ap_el1ns}\\
    H_{\{1,2\}} &=  - \frac{r_1^{2-\alpha}}{r_2^\beta} \frac{\Gamma (1-\alpha ) \Gamma (2-\alpha ) \Gamma (4-\alpha ) \Gamma (\alpha +1) \Gamma (1-\beta ) \Gamma (2-\beta ) \Gamma (\beta +1)  \Gamma \left(\tfrac{5}{2}-\alpha -\beta \right)}
    {\Gamma (3-\alpha ) \Gamma \left(\frac{9}{2}-\alpha \right) \Gamma \left(\frac{5}{2}-\beta \right)  \Gamma (5-\alpha -\beta)} \nonumber\\
    &\hspace{1cm}\times \Gamma (3-\alpha-\beta) \Gamma (\alpha +\beta +1)
    {}_5F_4 \left[ \begin{array}{c}
	    1, 2-\alpha, 4-\alpha, \tfrac{5}{2}-\alpha-\beta, 3-\alpha-\beta \\
    	3-\alpha, \tfrac{9}{2}-\alpha, 1-\alpha-\beta, 5-\alpha-\beta \\
    \end{array}	\bigg| r_1 \right]\\
    H_{\{1,3\}} &= - \frac{r_1^{3-\alpha}}{r_2^{\beta+1}} \frac{\Gamma (1-\alpha ) \Gamma (3-\alpha ) \Gamma (5-\alpha ) \Gamma (\alpha +1) \Gamma (1-\beta )^2 \Gamma (\beta +1)^2 \Gamma \left( \tfrac{5}{2}-\alpha -\beta \right)}
    {\Gamma (4-\alpha ) \Gamma \left(\frac{11}{2}-\alpha \right) \Gamma \left(\frac{3}{2}-\beta \right) \Gamma (\beta+2) \Gamma (5-\alpha-\beta)} \nonumber\\ 
   &\hspace{1cm} \times  \Gamma (3-\alpha -\beta) \Gamma (\alpha +\beta +1)
    {}_5F_4 \left[ \begin{array}{c}
	    1, 3-\alpha, 5-\alpha, \tfrac{5}{2}-\alpha-\beta, 3-\alpha-\beta \\
	    4-\alpha, \tfrac{11}{2}-\alpha, 1-\alpha-\beta, 5-\alpha-\beta \\
    \end{array}	\bigg| r_1 \right]
    \label{Ap_el3ns}\\
    H_{\{1,4\}} &= \frac{r_1^{4-\alpha}}{r_2^{2+\beta}}
    \frac{\Gamma (1-\alpha ) \Gamma (4-\alpha ) \Gamma (6-\alpha ) \Gamma (\alpha +1) \Gamma (1-\beta )^2 \Gamma (\beta +1) \Gamma (\beta +2) \Gamma \left(\tfrac{5}{2}-\alpha -\beta \right) }
    {\Gamma (5-\alpha ) \Gamma \left(\frac{13}{2}-\alpha \right) \Gamma \left(\frac{1}{2}-\beta \right) \Gamma (\beta +3) \Gamma (5-\alpha -\beta)} \nonumber\\
   &\hspace{-0.5cm} \times  \Gamma (3-\alpha -\beta) \Gamma (\alpha +\beta +1)
    F^{2:3;3}_{2:2;2} 
    \left[ \begin{array}{c}
	    4-\alpha, 6-\alpha: 1, \tfrac{1}{2}+\beta, 2+\beta; 1, \tfrac{5}{2}-\alpha-\beta, 3-\alpha-\beta\\
	    5-\alpha, \tfrac{13}{2}-\alpha: 1+\beta, 3+\beta; 1-\alpha-\beta, 5-\alpha-\beta  \\
    \end{array} \Bigg| \frac{r_1}{r_2},r_1 \right]
     \label{Ap_el4ns}
\end{align}

\bigskip

\setcounter{equation}{0}
\renewcommand{\theequation}{B-\arabic{equation}}

\subsection*{Muon case (see Table \ref{Table:MuonResults})}

\bigskip

\begin{align}
    h_{\{2,5\}} &= - \frac{r_1^{2-\alpha}}{r_2^\beta}
    \frac{\Gamma (1-\alpha ) \Gamma (2-\alpha ) \Gamma (4-\alpha ) \Gamma (1+\alpha) \Gamma (1-\beta ) \Gamma (2-\beta ) \Gamma (1+\beta) \Gamma \left(\tfrac{5}{2}-\alpha -\beta \right)}
    {\Gamma (3-\alpha ) \Gamma \left(\frac{9}{2}-\alpha \right) \Gamma \left(\frac{5}{2}-\beta \right) \Gamma (5-\alpha -\beta)} \nonumber \\
    &\hspace{2cm}\times  \Gamma (3-\alpha -\beta) \Gamma (1+\alpha +\beta)
    {}_5F_4 \left[ \begin{array}{c}
	1, 2-\alpha, 4-\alpha, 3-\alpha-\beta, \tfrac{5}{2}-\alpha-\beta \\
	3-\alpha, \tfrac{9}{2}-\alpha, 1-\alpha-\beta, 5-\alpha-\beta  \\
\end{array}	\bigg| r_1 \right] 
\label{Ap_mu1}\\
    h_{\{2,6\}} &= \frac{2}{\sqrt{\pi}} \frac{\Gamma \left(\frac{5}{2}-\alpha \right) \Gamma (3-\alpha)^2 \Gamma (\alpha +1)}
    {r_1^{\alpha-3} \; r_2 \; \Gamma (4-\alpha) \Gamma \left(\frac{11}{2}-\alpha \right)}
    {}_4F_3 \left[ \begin{array}{c}
	1, 3-\alpha, 3-\alpha, \tfrac{5}{2}-\alpha \\
	1-\alpha, 4-\alpha, \tfrac{11}{2}-\alpha  \\
\end{array}	\bigg| r_1 \right]\\
    h_{\{2,7\}} &=
    \frac{256 \; r_1^3 \;  \Gamma \left(\frac{3}{2}-\alpha \right)}
    {945\sqrt{\pi} \; r_2^{2+\alpha} \; \Gamma \left(\frac{1}{2}-\alpha \right) \Gamma (4-\alpha ) \Gamma (\alpha +3)} 
    {}_5F_4 \left[ \begin{array}{c}
	1, 3, 5, \tfrac{1}{2}+\alpha, 2+\alpha \\
	4, \tfrac{11}{2}, 1+\alpha, 3+\alpha \\
\end{array}	\bigg| \frac{r_1}{r_2} \right]\\
    h_{\{2,8\}} &=
    - \frac{16}{35\sqrt{\pi}} \frac{r_1^2}{r_2^{2+\alpha}} \frac{\Gamma (1-\alpha )^2 \Gamma (\alpha +2)^2}{ \Gamma (3-\alpha ) \Gamma (\alpha +3)}
    {}_5F_4 \left[ \begin{array}{c}
	1, 2, 4, \tfrac{1}{2}+\alpha, 2+\alpha \\
	3, \tfrac{9}{2}, 1+\alpha, 3+\alpha \\
\end{array}	\bigg| \frac{r_1}{r_2} \right]
\\
    h_{\{2,9\}} &= - \frac{8}{15\sqrt{\pi}} \frac{r_1}{r_2^{\alpha+2}} \frac{\Gamma \left(-\alpha -\frac{1}{2}\right) \Gamma (1-\alpha )^3 \Gamma (\alpha +1) \Gamma (\alpha +2)}
    {\Gamma \left(\frac{1}{2}-\alpha \right) \Gamma (2-\alpha)}
    {}_5F_4 \left[ \begin{array}{c}
	1, 1, 3, \tfrac{1}{2}+\alpha, 2+\alpha \\
	2, \tfrac{7}{2}, 1+\alpha, 3+\alpha \\
\end{array}	\bigg| \frac{r_1}{r_2} \right]
\\
    h_{\{2,10\}} &= \frac{8}{15 \sqrt{\pi}} \frac{r_1}{r_2^{\alpha+3}} \frac{ \Gamma \left(-\alpha -\frac{3}{2}\right) \Gamma (1-\alpha )^2 \Gamma (\alpha +1)^3 \Gamma (\alpha +3)}
    {\Gamma \left(-\alpha -\frac{1}{2}\right) \Gamma (\alpha +2)^2}
    {}_5F_4 \left[ \begin{array}{c}
	1, 1, 3, \tfrac{3}{2}+\alpha, 3+\alpha \\
	2, \tfrac{7}{2}, 2+\alpha, 4+\alpha \\
\end{array}	\bigg| \frac{r_1}{r_2} \right]
\\
    h_{\{2,11\}} &= \frac{r_1^{4-\alpha}}{r_2^{\beta+2}} \frac{\Gamma (1-\alpha ) \Gamma (4-\alpha ) \Gamma (6-\alpha ) \Gamma (\alpha +1) \Gamma (1-\beta )^2 \Gamma (\beta +1) \Gamma (\beta +2)\Gamma \left(\frac{5}{2}-\alpha -\beta \right)}
    { \Gamma (5-\alpha ) \Gamma \left(\frac{13}{2}-\alpha \right) \Gamma \left(\frac{1}{2}-\beta \right) \Gamma (\beta +3) \Gamma (5-\alpha -\beta)} \nonumber \\
    &{\hspace{-0.7cm}}\times  \Gamma (3-\alpha -\beta ) \Gamma (1+\alpha +\beta) 
    F^{2:3;3}_{2:2;2} \left[ \begin{array}{c}
	4-\alpha, 6-\alpha : 1, 2+\beta, \tfrac{1}{2}+\beta; 1, \tfrac{5}{2}-\alpha-\beta, 3-\alpha-\beta \\
	5-\alpha, \tfrac{13}{2}-\alpha : 1+\beta, 3+\beta; 1-\alpha-\beta, 5-\alpha-\beta
\end{array}	\bigg| \frac{r_1}{r_2}, r_1 \right]
\\
    h_{\{2,12\}} &= \frac{16}{15\sqrt{\pi}} \frac{r_1}{r_2^{4+\alpha}} \frac{\Gamma \left(-\alpha -\frac{5}{2}\right)\Gamma (\alpha +1) \Gamma (\alpha +4)}
    {\Gamma \left(-\alpha -\frac{3}{2}\right) \Gamma (\alpha +3)^2} \nonumber\\
    &\hspace{2cm}\times F^{2:3;3}_{2:2;2} \left[ \begin{array}{c}
	\tfrac{5}{2}+\alpha, 4+\alpha : 1, 1, 3; 1, 1+\alpha, 5+\alpha \\
    3+\alpha, 5+\alpha : 2, \tfrac{7}{2}; 3+\alpha, \tfrac{7}{2}+\alpha
\end{array}	\bigg| \frac{r_1}{r_2},\frac{1}{r_2} \right]
\label{Ap_mu8}
\end{align}

\newpage

\subsection*{Tau case (see Table \ref{Table:TauResults})}

\bigskip

\setcounter{equation}{0}
\renewcommand{\theequation}{C-\arabic{equation}}

\begin{align}
	h_{\{3,3\}} &= -\frac{8}{3 \sqrt{\pi}} \frac{1}{r_2^2} \bigg( -\log ^2(r_1) \log (r_2)-\frac{1}{3} \log (r_1) \log (r_2) -\frac{2}{3} \pi ^2 \log (r_1 r_2)+\frac{1}{3} \log ^3(r_1) \nonumber \\
	&\hspace{2.35cm} -\frac{13}{6} \log ^2(r_1) -\frac{25}{18} \log (r_1) - \frac{68}{9} \log (r_2) + 12 \zeta (3) - \frac{5 \pi ^2}{3}-\frac{301}{9} \bigg)
	\label{Ap_tau1}
\\
	h_{\{3,10\}} &= -\frac{8}{3\sqrt{\pi}} \frac{1}{r_1^2} \bigg( -\log ^2(r_1) \log (r_2)-\frac{14}{3} \log (r_1) \log (r_2)-\frac{2}{3} \pi ^2 \log (r_1 r_2) +\frac{1}{3} \log ^3(r_1) \nonumber \\
	&\hspace{2.35cm} +\frac{13}{6} \log ^2(r_1) + \frac{5}{6} \log (r_1) - \frac{88}{9} \log (r_2) + 12 \zeta (3)-\frac{5 \pi ^2}{3}-\frac{1847}{108} \bigg)
\\
    h_{\{3,13\}} &= - r_1^{-\alpha } r_2^{-\beta -3} \frac{ \Gamma (2-\alpha ) \Gamma (\beta +3) \Gamma \left(-\alpha -\beta -\frac{5}{2}\right)}{\Gamma \left(\frac{5}{2}-\alpha \right) \Gamma \left(-\beta -\frac{1}{2}\right) \Gamma (\beta +2) \Gamma (\beta +4) \Gamma (\alpha +\beta +3)} \nonumber \\
    & \times \Gamma (\alpha +\beta +1) \Gamma (\alpha +\beta +5)
    {}_5F_4 \left[ \begin{array}{c}     1,\tfrac{3}{2}+\beta,3+\beta,1+\alpha+\beta,5+\alpha+\beta \\
    2+\beta,4+\beta,3+\alpha+\beta,\tfrac{7}{2}+\alpha+\beta \\
    \end{array}	\bigg| \frac{1}{r_2} \right]
\\
    h_{\{3,14\}} &= \frac{45\pi}{8} \frac{r_1^{\alpha+\frac{1}{2}}}{4^\alpha r_2^{\alpha+2}} \frac{\Gamma \left(\alpha +\frac{1}{2}\right) \Gamma (\alpha +2) }
    {\Gamma (\alpha +1) \Gamma (\alpha +3)}
    {}_5F_4 \left[ \begin{array}{c}                 1,-\tfrac{1}{2},\tfrac{1}{2}+\alpha,\tfrac{7}{2},2+\alpha \\	    1+\alpha,\tfrac{3}{2},3+\alpha,2 \\
    \end{array}	\bigg| \frac{1}{r_2} \right]
\\
    h_{\{3,15\}} &= \frac{2}{\sqrt{\pi}} \frac{r_1^{1+\alpha}}{r_2^{2+\alpha}} \frac{\Gamma \left(\alpha +\frac{1}{2}\right)}
  {\Gamma \left(\alpha +\frac{7}{2}\right)}
  {}_2F_1 \left[ \begin{array}{c}
	    1, \tfrac{1}{2}+\alpha \\
	    \tfrac{7}{2}+\alpha \\
    \end{array}	\bigg| \frac{r_1}{r_2} \right]
\\
    h_{\{3,16\}} &= - \frac{4}{\sqrt{\pi}} \frac{r_1^{\alpha+1}}{r_2^{\alpha+3}} \frac{\Gamma (\alpha +1) \Gamma \left(\alpha +\frac{3}{2}\right) \Gamma (\alpha +3)^2}{\Gamma (\alpha +2)^2 \Gamma \left(\alpha +\frac{7}{2}\right) \Gamma (\alpha +4)}
    {}_5F_4 \left[ \begin{array}{c}
	    1, 1+\alpha, \tfrac{3}{2}+\alpha, 3+\alpha, 3+\alpha \\
	    2+\alpha, 2+\alpha, \tfrac{7}{2}+\alpha, 4+\alpha \\
    \end{array}	\bigg| \frac{r_1}{r_2} \right]
\\
    h_{\{3,17\}} &= \frac{35 \pi}{8} \frac{r_1^{\alpha -\frac{1}{2}}}{4^\alpha r_2^{\alpha+2}} \frac{\Gamma \left(\alpha +\frac{1}{2}\right) \Gamma (\alpha +2)}{\Gamma (\alpha +1) \Gamma (\alpha+3)}
    {}_5F_4 \left[ \begin{array}{c}
	    1, \tfrac{1}{2}+\alpha, \tfrac{1}{2}, 2+\alpha, \tfrac{9}{2} \\
	    1+\alpha, \tfrac{5}{2}, 3+\alpha, 3 \\
    \end{array}	\bigg| \frac{1}{r_2} \right]
\\
    h_{\{3,18\}} &= - \frac{2}{\sqrt{\pi} r_1 r_2^{\alpha+2}} \frac{\Gamma \left(\alpha +\frac{1}{2}\right) \Gamma (\alpha +2) \Gamma (\alpha +5) }{\Gamma (\alpha +3)^2 \Gamma \left(\alpha +\frac{7}{2}\right)}
    {}_4F_3 \left[ \begin{array}{c}
	    1, \tfrac{1}{2}+\alpha, 2+\alpha, 5+\alpha \\
	    3+\alpha, 3+\alpha, \tfrac{7}{2}+\alpha \\
    \end{array}	\bigg| \frac{1}{r_2} \right]
\\
    h_{\{3,19\}} &= \frac{7\pi}{8} \frac{r_1^{\alpha -\frac{3}{2}}}{4^\alpha r_2^{\alpha+2}}
    \frac{ \Gamma \left(\alpha +\tfrac{1}{2}\right) \Gamma (\alpha +2)}
    {\Gamma (\alpha +1) \Gamma (\alpha +3)}
    {}_5F_4 \left[ \begin{array}{c}             1,\tfrac{3}{2},\tfrac{11}{2},\tfrac{1}{2}+\alpha,2+\alpha \\	    \tfrac{7}{2},4,1+\alpha,3+\alpha \\ \end{array}	\bigg| \frac{1}{r_2} \right]
\\
    h_{\{3,20\}} &= \frac{2}{\sqrt{\pi}} \frac{1}{r_2 r_1^{\alpha+2}}
    \frac{\Gamma \left(-\alpha -\frac{5}{2}\right) \Gamma (\alpha +2) \Gamma (\alpha +5)}
    {\Gamma \left(\frac{1}{2}-\alpha \right) \Gamma (\alpha +3)^2}
    {}_4F_3 \left[ \begin{array}{c}
	    1,2+\alpha,5+\alpha,\tfrac{1}{2}+\alpha \\
	    3+\alpha,3+\alpha,\tfrac{7}{2}+\alpha \\
    \end{array}	\bigg| \frac{1}{r_1} \right]
\\
    h_{\{3,21\}} &= \frac{ 14 \pi^{3/2}}{3 r_1^{5/2}}
    {}_4F_3 \left[ \begin{array}{c}
	    1,1,\tfrac{1}{2},\tfrac{9}{2} \\
	    \tfrac{3}{2},3,\tfrac{7}{2} \\
    \end{array}	\bigg| \frac{1}{r_1} \right]
\end{align}

\begin{align}
    h_{\{3,22\}} &= - r_1^{-\alpha -3} r_2^{-\beta } \frac{ \Gamma (\alpha +3)  \Gamma (2-\beta )  \Gamma \left(-\alpha -\beta -\frac{5}{2}\right)}{\Gamma \left(-\alpha -\frac{1}{2}\right) \Gamma (\alpha +2) \Gamma (\alpha +4) \Gamma \left(\frac{5}{2}-\beta \right) \Gamma (\alpha +\beta +3)} \nonumber \\
    &\hspace{0.1cm}\times \Gamma (\alpha +\beta +1) \Gamma (\alpha +\beta +5) 
    {}_5F_4 \left[ \begin{array}{c}
	    1, \tfrac{3}{2}+\alpha, 3+\alpha, 1+\alpha+\beta, 5+\alpha+\beta \\
	    2+\alpha, 4+\alpha, 3+\alpha+\beta, \tfrac{7}{2}+\alpha+\beta \\
    \end{array}	\bigg| \frac{1}{r_1} \right]
\\
    h_{\{3,23\}} &= - \pi^{3/2} \frac{r_1^{\alpha +\tfrac{3}{2}}}{r_2^{\alpha+2}} \frac{\Gamma \left(\alpha +\tfrac{1}{2}\right) \Gamma \left(\alpha +\tfrac{3}{2}\right) \Gamma (\alpha +2) \Gamma \left(\alpha +\tfrac{7}{2}\right)}
    {\Gamma \left(\tfrac{5}{2}-\alpha \right) \Gamma \left(\alpha -\tfrac{3}{2}\right) \Gamma (\alpha +1) \Gamma \left(\alpha +\tfrac{5}{2}\right) \Gamma (\alpha +3) \Gamma (\alpha +4)} \nonumber \\
   &\hspace{2cm} \times 
    F^{2:3;2}_{2:2;1} \left[ \begin{array}{c}
    	\tfrac{1}{2}+\alpha, 2+\alpha: 1, \tfrac{3}{2}+\alpha, \tfrac{7}{2}+\alpha;  -\tfrac{3}{2}, \tfrac{5}{2} \\
        1+\alpha, 3+\alpha: \tfrac{5}{2}+\alpha, 4+\alpha; \tfrac{1}{2}
    \end{array}	\bigg| \frac{r_1}{r_2}, \frac{1}{r_2} \right]
\\
    h_{\{3,24\}} &= \frac{r_1^{-\alpha -2}}{r_2^{\beta+2}} \frac{\Gamma (1-\alpha )  \Gamma (\alpha +2) \Gamma (1-\beta )  \Gamma (\beta +2) \Gamma \left(-\alpha -\beta -\frac{7}{2}\right)  \Gamma (\alpha +\beta +2)\Gamma (\alpha +\beta +6)}
    {\Gamma \left(\frac{1}{2}-\alpha \right) \Gamma (\alpha +3) \Gamma \left(\frac{1}{2}-\beta \right) \Gamma (\beta +3) \Gamma (\alpha +\beta +4)} \nonumber \\
    &\hspace{2cm}\times 
    F^{2:3;3}_{2:2;2} \left[ \begin{array}{c}
    	2+\alpha+\beta, 6+\alpha+\beta: 1,\tfrac{1}{2}+\alpha, 2+\alpha; 1, \tfrac{1}{2}+\beta, 2+\beta \\
        4+\alpha+\beta, \tfrac{9}{2}+\alpha+\beta: 1+\alpha, 3+\alpha; 1+\beta, 3+\beta
\end{array}	\bigg| \frac{1}{r_1}, \frac{1}{r_2} \right]
\\
    h_{\{3,25\}} &= - \frac{99 \pi ^{3/2}}{320 r_1^{5/2} r_2^2} 
    F^{2:3;2}_{2:2;1} \left[ \begin{array}{c}
    	\tfrac{5}{2}, \tfrac{13}{2}: 1,1,\tfrac{5}{2}; \tfrac{1}{2},2 \\
        \tfrac{9}{2}, 5: \tfrac{3}{2}, \tfrac{7}{2}; 3
\end{array}	\bigg| \frac{1}{r_1}, \frac{1}{r_2} \right]
\label{Ap_tau15}
\end{align}

\end{document}